\documentclass[10pt,twocolumn,letterpaper]{article}

\usepackage{iccv}
\usepackage{times}
\usepackage{epsfig}
\usepackage{graphicx}
\usepackage{amsmath}
\usepackage{amssymb}
\usepackage{amsthm}
\newtheorem{definition}{\protect\defname}
\newtheorem{prop}{\protect\propositionname}

\newtheorem{thm}{\protect\theoremname}
\newtheorem{lem}{\protect\lemmaname}
\providecommand{\lemmaname}{Lemma}
\providecommand{\defname}{Definition}
\providecommand{\propositionname}{Proposition}
\providecommand{\remarkname}{Remark}
\providecommand{\theoremname}{Theorem}
\DeclareMathOperator*{\argmin}{arg\,min}

\usepackage{caption}
\usepackage{subcaption}
\usepackage{dirtytalk}

\usepackage[accsupp]{axessibility}

\usepackage[pagebackref=true,breaklinks=true,letterpaper=true,colorlinks,bookmarks=false]{hyperref}

\iccvfinalcopy 

\def\iccvPaperID{11087} 

\ificcvfinal\fi

\begin{document}

\title{Shape Analysis of Euclidean Curves under Frenet-Serret Framework}

\author{Perrine Chassat$^1$ \\
$^1$LaMME, University of Paris-Saclay\\
{\tt\small perrine.chassat@univ-evry.fr}
\and
Juhyun Park$^{1,2}$ \\
$^2$ENSIIE, Evry \\
{\tt\small juhyun.park@ensiie.fr}
\and
Nicolas Brunel$^{1,2,3}$ \\
$^3$Quantmetry, Paris \\
{\tt\small nicolas.brunel@ensiie.fr}
}

\maketitle
\ificcvfinal\thispagestyle{empty}\fi

\begin{abstract}
Geometric frameworks for analyzing curves are common in applications as they focus on invariant features and provide visually satisfying solutions to standard problems such as computing invariant distances, averaging curves, or registering curves.
We show that for any smooth curve in $\mathbb{R}^d, d>1$, the generalized curvatures associated with the Frenet-Serret equation can be used to define a Riemannian geometry that takes into account all the geometric features of the shape.
This geometry is based on a Square Root Curvature Transform that extends the square root-velocity transform for Euclidean curves (in any dimensions) and provides likely geodesics that avoid artefacts encountered by representations using only first-order geometric information. Our analysis is supported by simulated data and is especially relevant for analyzing human motions. We consider trajectories acquired from sign language, and show the interest of considering curvature and also torsion in their analysis, both being physically meaningful.

\end{abstract}

\section{Introduction}

Identifying and comparing different types of visual objects is a fundamental task in machine learning and computer vision problems \cite{LRSBC16, eisenberger2020hamiltonian, koestler2022intrinsic}.
The shape is one of the essential features of objects that allow us to understand and characterize them. Nowadays, it is much easier to obtain data in the form of shapes, typically as dense point clouds or landmarks. 
The main task in shape analysis is to define a proper framework to compare and quantify the variation of the shapes. 
However, the shape space is generally nonlinear, and extracting meaningful information or features is complex.
One of the successful approaches to shape analysis utilizes a Riemannian framework of differential geometry,  
where a metric can be defined between the shapes, which is invariant with respect to shape-preserving transformations such as translation and rotation. For instance, this gives rise to geodesic distances that are naturally invariant to smooth and optimal deformations through geodesic paths between the shapes \cite{Lang2006}. 
%
This approach is very versatile as it can be adapted to various kinds of manifold-value data and can be designed to emphasize important geometric information to be preserved. As a consequence, several choices of metrics are possible, such as the class of invariant Sobolev metrics, often called \textit{elastic}, for the analysis of curves \cite{Bauer2021}.
In this work, we are concerned with curves that often arise in the application as trajectories (function of time) or motions (animation, activity recognition) and with the definition of a framework to compare their shapes. The differential geometry of Euclidean curves is among the simplest (with respect to higher dimensional manifolds), and relatively simple Riemannian metrics are available with different mathematical representations of curves \cite{Younes1998}. 
Quite remarkably, the introduction of the Square Root Velocity (SRV) transform \cite{Srivastava2011} that consists of a particular representation of the shape of a curve enables to define a so-called elastic Riemannian distance, which has proven to be useful for the statistical shape analysis of 2D and 3D curves in applications. The SRV possesses interesting properties such as a principled theoretical framework, efficient computation, and generalization to higher dimensions \cite{Bauer2022}.
Nevertheless, a limitation of the SRV transform and the corresponding elastic distance is the restrictive use of the first-order derivative, while the geometry of 3-D (or $d$-D) curves depends on the derivatives until order $d$. Indeed, it is well-known that a 3D curve is characterized by its curvature and its torsion: this is particularly critical when we consider trajectories or human movements, where the curvature and torsion can have a physical meaning.

\section{Related works and contributions}

As we will recall in section \ref{sec:Background}, the full geometry of a curve can be given either by the Frenet curvatures (standard curvature and torsion in 3D) or by the path of Frenet frames. There have been few attempts to directly deal with the Frenet curvatures: most of the works have been produced in 2D curves as an alternative representation \cite{BookFunctionalShapeDataAnalysis}. Nevertheless, the potential for applications has not been investigated. In \cite{Needham2019}, the elastic shape analysis framework has been considered for 3D curves based on the Frenet frames, but the link to the physical parameters has been overlooked.
We can also mention the shape analysis of curves on Lie groups \cite{Celledoni2016} with application in computer animation. Outside the Riemannian framework, an attempt has been made to use a direct curvature-based interpolation of curves \cite{Saba2012, Surazhsky2002}.

\begin{figure*}
\begin{center}
\includegraphics[width=0.83\linewidth]{ 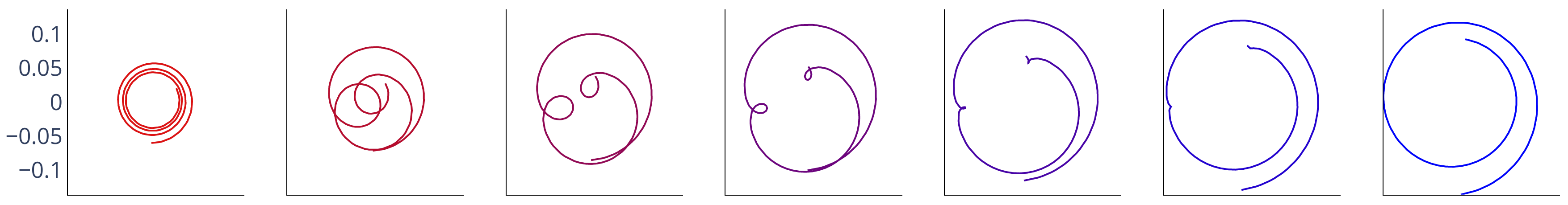}
\includegraphics[width=0.83\linewidth]{ 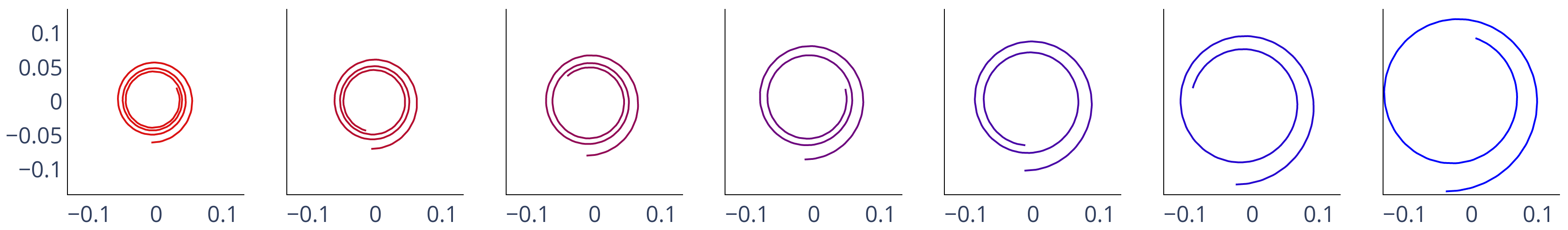}
\end{center}
   \caption{Geodesic paths between two 2D scaled spirals with different number of spins: SRVF ($1^{st}$ row) and SRC ($2^{nd}$ row).}
\label{fig:geod_spiral}
\end{figure*}

In this work, we introduce two representations of Euclidean curves for their shape analysis that use their complete geometry through Frenet curvatures. We provide the full development of the Riemannian frameworks associated with these two representations. As a consequence and in comparison with existing methods, our approaches also give explicit formulas for geodesics and geodesic distances. The first representation considered is based directly on unparametrized Frenet curvatures. We show through experiments that it defines a shape analysis framework that lacks elasticity. As the main contribution, we propose the definition of a second representation, called the Square-Root Curvature (SRC) Transform, which takes into account reparameterization and defines a metric on the space of shapes through the quotient space with the group of diffeomorphisms. One can imagine that the classical method associated with the SRVF, defining a Sobolev elastic metric \cite{Bauer2022}, already implicitly uses all the geometric information necessary for a relevant curve analysis. We show here with simple examples that this is not the case. Through experiments on synthetic data, 
we compare the methods, and illustrate the limitations of the SRVF one, due to its lack of use of geometric information. To be able to judge and compare the quality of these metrics, we compare consistent sets of curves characterized by specific features. The SRC method shows a special strength in defining a framework that remains consistent with these sets. The straightforward example of geodesics between helices with different numbers of spins (Figure~\ref{fig:geod_spiral} in 2D and Figure~\ref{fig:geod_x_helix} in 3D) shows that, in contrast, this is not the case for the SRVF method. In addition, we highlight the interest of these Frenet curvatures-based representations in the real application case of human motion trajectory analysis.

\section{Riemannian Geometry on Shape Space \label{sec:Background}}

We introduce useful notations and we review the main approach for constructing tractable representations of the shape of a curve and deriving a Riemannian geometry.

\subsection{Shape Analysis of Euclidean Curves}

We consider absolutely continuous curves that are smooth, open, and with values in some Euclidean space $\mathbb{R}^d$, we denote this set as $AC\left([0,1],\mathbb{R}^d \right)$. These curves are typically parametrized by a variable $t$ that can usually be interpreted as time.  
Nevertheless, from a (statistical) shape analysis point of view,  we focus on the \textit{geometric shape} of curves that do not depend on a specific parametrization or standard transformations  
such as translations, rotations, scaling, or reparametrizations. To distinguish between parametrized curves that differ only by translation, we consider the set of absolutely continuous curves where $x(0) = 0$, denoted by $AC_0\left([0,1],\mathbb{R}^d \right)$. The natural and intrinsic parametrization that uniquely defines the \textit{shape} of a curve $x$ is the arc-length parametrization, defined with the arc length function $s(t) = \int_0^t \|\dot{x}(u)\|du$, for $t\in[0,1]$. In order to remove the scaling variability, the total length of the curve $s(1)$ is set to 1.
Under this parametrization, the \textit{shape} $X: [0,1] \mapsto \mathbb{R}^d$ of the curve is the image of the function $x$ such that $x(t) = X(s(t))$. As we want to study shapes independently of their parameterizations, we introduce the reparametrization group $\textrm{Diff}_+([0,1])$, of smooth orientation preserving diffeomorphisms of the interval $[0,1]$ onto itself. This group acts on the space of absolutely continuous curves by right composition, and this action only alters the parametrization of the curve, not the inherent shape $X$. 
The space of such shapes (or unparametrized curves) is often mathematically defined as the quotient space
\begin{equation}
    \mathcal{S}([0,1], \mathbb{R}^d) = AC_0\left([0,1],\mathbb{R}^d \right)/\textrm{Diff}_+([0,1]).
\end{equation}
The purpose of shape analysis of curves is to define a distance function $d_\mathcal{S}$ on $\mathcal{S}$ and a framework to perform a complete statistical analysis on a set of curves in $\mathcal{S}$ (\eg mean, classification, or Principal Component Analysis etc.). One of the main challenges in defining this distance is to choose an appropriate mathematical representation of the curves that can be made invariant to all shape-preserving transformations - translation, rotation, scaling, and reparametrization.
Moreover, one of the stakes of such representation is to offer an (infinite-dimensional) Riemannian manifold structure that brings powerful and flexible tools for studying the geometry of shapes or statistical properties notably thanks to the tangent space of the manifold \cite{Lang2006, Sommer2020}.  
In \cite{BookFunctionalShapeDataAnalysis}, a list of the few possible representations is given - coordinate functions, curvatures, angle function, and square-root velocity function (SRVF) - and a framework for curve analysis is derived for the last two ones.
While the angle representation is unparameterized, the SRVF representation depends on the parametrization, which is shown to be very useful as a tool for the registration of points across curves.
As a consequence, the parameterization group $\textrm{Diff}_+([0,1])$ must be eliminated by using a quotient space. The classical approach is to define the Riemannian metric on the shape space through a metric on the space of parametrized representations that is invariant to reparametrization: $\forall h \in \textrm{Diff}_+([0,1])$ \begin{equation}
d_{AC_0}(x_0,x_1) = d_{AC_0}(x_0 \circ h, x_1 \circ h). 
\end{equation}
In that case, the distance on $\mathcal{S}$ is defined as the infimum over all possible reparametrization. For $X_0, X_1 \in \mathcal{S}$, \begin{equation}
d_{\mathcal{S}}(X_0, X_1) := \inf_{h \in \textrm{Diff}([0,1])} d_{AC_0}(x_0, x_1 \circ h).
\end{equation}
In the following, we will denote with a dot the derivation with respect to the time variable, and with a prime the one with respect to the arc-length parameter.


\begin{figure*}
\begin{center}
\includegraphics[width=0.8\linewidth]{ 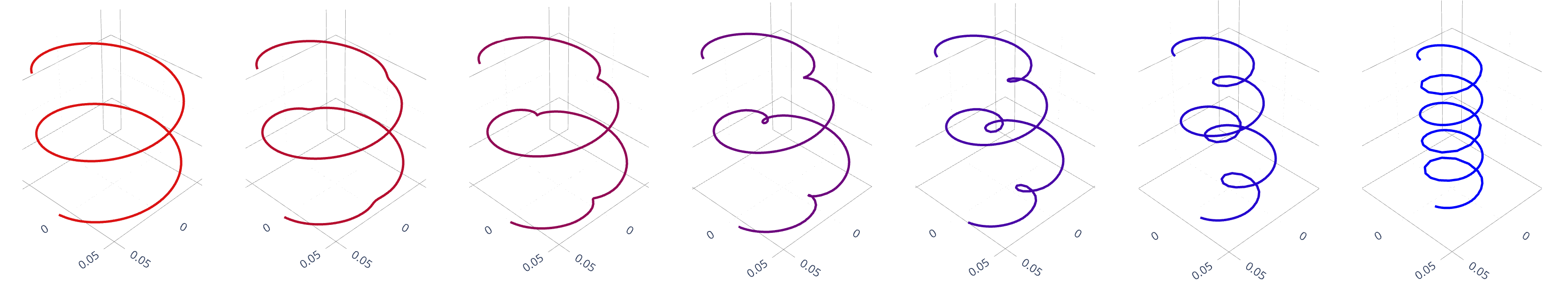}
\includegraphics[width=0.8\linewidth]{ 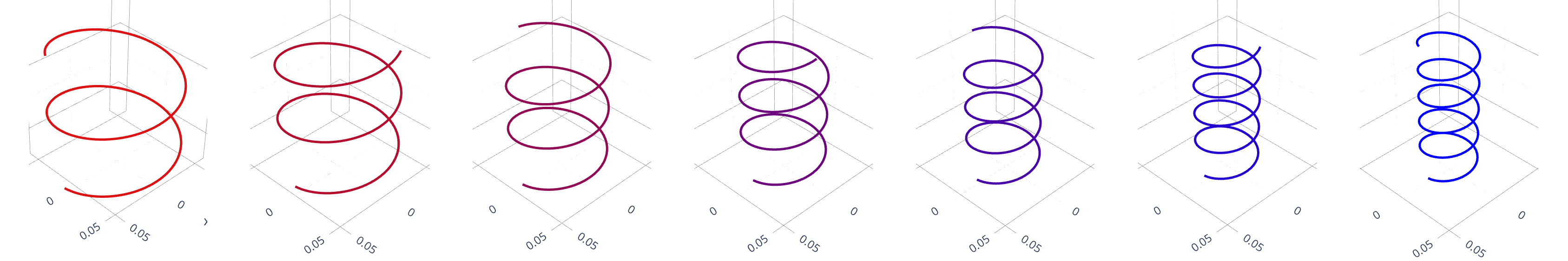}
\end{center}
   \caption{Geodesic paths between two scaled 3D circular helices with different number of spins: SRVF ($1^{st}$ row) and SRC ($2^{nd}$ row).}
\label{fig:geod_x_helix}
\end{figure*}

\subsection{Square Root Velocity Framework}

The square-root velocity function framework is the most commonly used representation for curve shape analysis in $\mathbb{R}^d$ \cite{Srivastava2011, BookFunctionalShapeDataAnalysis}. The square-root velocity function (SRVF) of  $x \in AC_0\left([0,1],\mathbb{R}^d \right)$, denoted by $\mathcal{R}_{\textrm{SRVF}}(x) = q$, is defined as
\begin{equation}
     q(t) = \frac{\dot{x}(t)}{\sqrt{\|\dot{x}(t)\|}} = \sqrt{\dot{s}(t)}T(s(t))
\label{eq:srvf_def}
\end{equation}
where $T(s(t)) = \frac{\dot{x}(t)}{\|\dot{x}(t)\|} = X'(s(t))$ is the unit tangent vector of the curve. 
This transformation is a bijection with $AC_0\left([0,1],\mathbb{R}^d \right)$ and its explicit inverse is $x(t) = \int_0^t q(u)|q(u)|du$.
As we consider length-normalized curves, the SRVFs have a unit $\mathbb{L}^2$ norm, and their set is the convenient unit Hilbert sphere, a Riemannian submanifold of $\mathbb{L}^2([0,1],\mathbb{R}^d)$ (with the $\mathbb{L}^2$ inner product). 
Then, the $\mathbb{L}^2$ metric on SRVF induces a Riemannian metric on $AC_0\left([0,1],\mathbb{R}^d \right)$ where geodesics are given by the shorter arcs on great circles between SRV functions. 
The action of $\textrm{Diff}_+([0,1])$ on $AC_0\left([0,1],\mathbb{R}^d \right)$ is reflected on $q$ by the group action denoted by $*$ and defined as
\begin{equation}
(q * h)(t) = \sqrt{\dot{h}(t)}q(h(t))
\label{eq:group_action_srvf}
\end{equation}
and if the curve is rotated by a matrix $O \in SO(d)$, its SRVF gets rotated by the same matrix. The key property of this representation is the invariance of its associated distance under the action of $\textrm{Diff}_+([0,1])$ and $SO(d)$:
\begin{equation}
\|O(q_0 * h) - O(q_1 * h)\|_{\mathbb{L}^2} = \|q_0 - q_1\|_{\mathbb{L}^2}.
\end{equation}
The metric can then be used to define a proper distance on the shape space $\mathcal{S}([0,1], \mathbb{R}^d)$
\begin{align}
    d_{\mathcal{S}}^{(\textrm{SRVF})}(X_0, X_1) := \inf_{\substack{O \in SO(d) \\ h \in \textrm{Diff}([0,1])}} \cos^{-1}  \langle q_0, O(q_1 * h) \rangle 
\end{align}
and the geodesic path on the shape space is taken between $q_0$ and $q_1*h$. 

The definition of this distance on the shape space under the SRVF representation can be interpreted as the following registration problem
\begin{equation}
h^*, O^* = \argmin_{\substack{O \in SO(d) \\ h \in \textrm{Diff}([0,1])}} \int_0^1 \|q_0(t) - Oq_1(h(t))\sqrt{\dot{h}(t)}\|^2_{2}dt\,.
\label{eq:reg_pb_srvf_h}
\end{equation}
In \cite{Brunel2019}, this registration problem has been reformulated with the unit tangent vector and the arc length functions. By defining $\gamma = s_1 \circ h \circ s_0^{-1} \in \textrm{Diff}_+([0,1])$ the optimization problem \ref{eq:reg_pb_srvf_h} amounts to finding the optimal diffeomorphism of $\textrm{Diff}_+([0,1])$ that acts on the arc-length parameter $s$ and solves the minimization problem:
\begin{equation}
\gamma^*, O^* = \argmin_{\substack{O \in SO(d) \\ \gamma \in \textrm{Diff}([0,1])}} \int_0^1 \|T_0(s) - OT_1(\gamma(s))\sqrt{\gamma'(s)}\|^2_{2}ds \,.
\label{eq:reg_pb_srvf_T}
\end{equation}
It should be noted, in this reformulation, that the object $OT_1(\gamma(s))\sqrt{\gamma'(s)}$ does not represent the same shape as $X_1(s)$ in the shape space. Here the element $\gamma$ of $\textrm{Diff}_+([0,1])$ is not used as a reparametrization of the curve but to deform the element of $\mathcal{S}([0,1], \mathbb{R}^d)$. 
Under this point of view, the set of unit tangent vectors that can be reached by deforming the vector $T(s)$, with the group action $T * \gamma$ defines an equivalence class of shapes associated with that one, as in the setting of deformable templates of Grenander’s theory \cite{Younes2000, Younes2018}. 

Finally, the choice of a parametrized curve representation for shape analysis, discussed in \cite{BookFunctionalShapeDataAnalysis}, can be seen as the problem of choosing a good geometric representative of the shape as a template and defining an associated registration problem. 
Hence, an appropriate choice may be seen as a matter of modeling and should be done in interaction with the type of data analyzed and the dimension of the space. In the next sections, we use $h$ to refer to functions of $\textrm{Diff}_+([0,1])$ that act on the time variable $t$ and $\gamma$ for ones that act on the arc-length variable $s$.


\section{Exhaustive geometric information with Frenet representation}

Based on these previous observations and with the intention of developing a more suitable framework for the analysis of three-dimensional curves, Brunel and Park \cite{Brunel2019} proposed a direct extension of the SRVF method, which considers not only the tangent vector as the geometric representation of the curve but the whole Frenet-Serret frame in three dimensions. Their idea is to use an exhaustive description of the geometry of curves by incorporating higher-order information about the geometry in the representation. To exploit this idea, we propose to study suitable representations based on this Frenet-Serret framework. Proofs of theoretical results are given in the supplementary material.

\subsection{The Frenet-Serret framework}

 We introduce the Frenet-Serret framework for curves of any dimension $d$. Let $F([0,1],\mathbb{R}^d)$ be the set of curves $x \in AC_0\left([0,1],\mathbb{R}^d \right)$ $d$-times continuously diﬀerentiable, and with the first $d$ derivatives linearly independent. $F([0,1],\mathbb{R}^d)$ is called the set of \textit{Frenet curves}. In the following, we will restrict the shape space to be the set 
\begin{align}
\mathcal{S}([0,1],\mathbb{R}^d) = F([0,1],\mathbb{R}^d)/\textrm{Diff}_+([0,1]) \,.
\end{align}
The \textit{Frenet frame} $e_1, e_2, \hdots, e_d$ associated with $X \in \mathcal{S}([0,1],\mathbb{R}^d)$ is uniquely defined by applying the Gram-Schmidt process to the first $d$ derivatives of $X$.
In dimension $3$, the three vectors of the Frenet frame are known as the tangent, normal and bi-normal vector. We define the function $Q$ that maps to $s \in [0,1]$ along the curve the corresponding Frenet frame
\begin{equation}
Q(s)= \left[e_1(s) \, | \, e_2(s) \, | \, ... \, |\,  e_d(s) \right].
\end{equation} 
 The function $Q$ is a measurable curve from $[0,1]$ to the Lie group of rotation matrices $SO(d)$ called the \textit{Frenet path}. 
 \begin{thm}[Frenet-Serret equation \cite{Wolfgang}] \label{th:Frenet_ODE}
Let $X \in \mathcal{S}([0,1],\mathbb{R}^d)$ and $Q(s)$ its associated Frenet path. Then there are functions $\theta_1 , \hdots, \theta_{d-1}$ deﬁned on that curve with $\theta_1 , \hdots, \theta_{d-2} > 0$, so that every $\theta_i$ is $(d-1-i)$-times continuously diﬀerentiable and
\begin{equation}
Q'(s) = Q(s) A_{\boldsymbol{\theta}}(s)
\end{equation}
where $\boldsymbol{\theta}(s) = (\theta_1(s), \hdots, \theta_{d-1}(s))^T$ and
\begin{equation*} 
A_{\boldsymbol{\theta}}(s) = \begin{footnotesize}\begin{bmatrix}
            0 & -\theta_1(s) & 0 & \hdots  & 0 \\
            \theta_1(s) & 0 & - \theta_2(s) & \ddots & \vdots   \\
            0 & \theta_2(s) & \ddots  & \ddots & 0 \\
            \vdots & \ddots & \ddots & 0 & -\theta_{d-1}(s) \\
            0 & \hdots & 0 & \theta_{d-1}(s) & 0
            \end{bmatrix} \end{footnotesize}
\end{equation*}
and $\theta_i$ is called the $i$-th Frenet curvature, and the equation is called the Frenet-Serret equation.
\end{thm}
In dimension 3, the two Frenet curvatures are known as $s \mapsto \kappa(s)$ the curvature function and $s \mapsto \tau(s)$ the torsion function. They have an interpretable physical meaning. The curvature function measures how sharply the curve changes direction at a given point, and the torsion function measures the degree to which the curve twists and turns as it moves along its path. The Frenet-Serret equation with an initial condition $Q(0)=Q_0$ defines an ordinary differential equation on the Lie group $SO(d)$ where the function $s \mapsto A_{\boldsymbol{\theta}}(s)$ has values in the Lie algebra of skew-symmetric matrices. This equation can also be expressed in function of the time variable $t$ as
\begin{equation}
\frac{dQ(s(t))}{dt} = \dot{s}(t)Q(s(t))A_{\boldsymbol{\theta}}(s(t)) \,.
\end{equation}
\begin{lem}
The Frenet curvatures and the Frenet path are invariant under all Euclidean motions.
\label{lem:invariance_curv_Q}
\end{lem}
This means that for $x \in F([0,1],\mathbb{R}^d), O \in SO(d), a \in \mathbb{R}^d$ and $h \in \textrm{Diff}_+ ([0,1])$, the curve defined by $\Tilde{x}(t) = a + Ox(h(t))$ has the same Frenet curvatures as $x$ and $\Tilde{Q}(s) = OQ(s)$. 
\begin{thm}[Fundamental theorem of the local theory of curves \cite{Wolfgang}]
Let $\theta_1,...,\theta_{d-1} \in C^{\infty}([0,1],\mathbb{R})$ such that $\theta_1, \hdots, \theta_{d-2} > 0$. For a given $X_0 \in \mathbb{R}^d$ and $Q_0 \in SO(d)$ there is a unique $X \in \mathcal{S}([0,1],\mathbb{R}^d)$ parametrized by arc length and satisfying the following three conditions:
\begin{itemize}
    \item $X(0) = X_0$,
    \item $Q_0$ is the Frenet frame of $X$ at point $s = 0$,
    \item $\theta_1,...,\theta_{d-1}$ are the Frenet curvatures of $X$. 
\end{itemize}
\label{th:Fund_th_curves}
\end{thm}
Theorems \ref{th:Frenet_ODE} and \ref{th:Fund_th_curves} state that there is a bijection, up to a translation and a rotation, between the shape space $\mathcal{S}([0,1],\mathbb{R}^d)$, the set of admissible Frenet curvatures $\mathcal{H}$ and the set of corresponding Frenet paths $\mathcal{F}_0$, where 
\begin{align}
\mathcal{H} = \left \{ \boldsymbol{\theta} \in C^{\infty}([0,1],\mathbb{R}^{d-1}) | \theta_1, \hdots, \theta_{d-2} > 0 \right \}\,,
\end{align}
\begin{equation}
\mathcal{F}_0 = \left \{ \begin{array}{c}
    Q \in \mathbb{L}^2([0,1], SO(d)) \textrm{ such that} \\
    Q'(s) = Q(s)A_{\boldsymbol{\theta}}(s),  Q(0) = I_d , \boldsymbol{\theta} \in \mathcal{H}
\end{array} \right \}\,.
\end{equation} 

\vspace{0.2cm}
From the detailed Frenet-Serret framework, one can think of the direct extension of the square-root velocity function \ref{eq:srvf_def} that simply consists in replacing the tangent vector with the entire Frenet frame. The representation of a parametrized curve $x \in F([0,1],\mathbb{R}^d)$ will be then
\begin{equation}
\mathcal{R}_{Q}(x)(t) = \sqrt{\dot{s}(t)}Q(s(t)).
\end{equation}
This representation is used in \cite{Brunel2019} to define a new alignment method on $\mathcal{S}$. They extend the SRVF registration problem \ref{eq:reg_pb_srvf_T} by using the Frobenius distance between the Frenet frames instead of only the $\mathbb{L}^2$ distance of the unit tangent vectors. They show to obtain more precise results with their method than the SRVF one. From the previous theorems, it is clear that this representation uniquely defines a parametrized curve $x \in F([0,1],\mathbb{R}^d)$. 
However, as a result of the Frobenius theorem \cite{Lyons2016}, the set of such representations $\mathcal{R}_Q$ appears not to be a manifold. We leave the demonstration in the supplementary material.



\subsection{Unparametrized Frenet curvatures}

A possible representation of a parametrized curve, already suggested in \cite{BookFunctionalShapeDataAnalysis, Surazhsky2002, Saba2012}, which keeps the idea of encoding more geometric information, is the unparametrized Frenet curvatures and the arc-length function pair 
\begin{equation}
    \mathcal{R}_{\boldsymbol{\theta}}(x)(t) = \left(\sqrt{\dot{s}(t)}, \, \boldsymbol{\theta}(s(t)) \right).
\end{equation}
We denote $\Psi([0,1])$, the set of square root velocity functions of length-normalized arc-length functions. This set is well-studied in the literature \cite{Marron2015, Tucker2013}. It is the unit sphere of the Hilbert space $\mathbb{L}^2([0,1],\mathbb{R})$ and therefore a Riemannian manifold equipped with the $\mathbb{L}^2$ metric. Then, the geodesic distance between two elements in $\Psi([0,1])$ is 
\begin{equation}
d_{\Psi}\left(\sqrt{\dot{s}_0}, \sqrt{\dot{s}_1} \right) = \cos^{-1}\left( \left \langle \sqrt{\dot{s}_0}, \sqrt{\dot{s}_1}\right \rangle \right)
\end{equation}
and the geodesic path connecting them is given by
\begin{equation}
\alpha_\Psi(\tau) = \frac{\sin((1-\tau)\vartheta)}{\sin(\vartheta)} \sqrt{\dot{s}_0} + \frac{\sin(\tau \vartheta)}{\sin(\vartheta)} \sqrt{\dot{s}_1}
\end{equation}
where $\vartheta = d_{\Psi}(\sqrt{\dot{s}_0}, \sqrt{\dot{s}_1})$. 
Moreover, any element of $\textrm{Diff}_+([0,1])$ is uniquely represented by an element of $\Psi([0,1])$.
\begin{prop}
The set of Frenet curvatures $\mathcal{H}$ is a Riemannian submanifold of $\mathbb{L}^2([0,1],\mathbb{R}^{d-1})$. 
\end{prop}
\begin{proof}
The set $M=\{ x \in \mathbb{R}^{d-1} | x_1,\hdots,x_{d-2} > 0 \}$ is a open subset of the Riemannian manifold $\mathbb{R}^{d-1}$. Then it is itself a differentiable Riemannian manifold with the standard inner product of $\mathbb{R}^{d-1}$, and for any point $p \in M$ the tangent space $T_p (M)$ is $\mathbb{R}^{d-1}$. The set of Frenet curvatures $\mathcal{H}$ is the set of measurable curves from $[0,1]$ to the Riemannian manifold $M$ and thus also a manifold (\cite{Wang1989}). Its tangent space is $\mathbb{L}^2([0,1],\mathbb{R}^{d-1})$ and it can be equipped with the $\mathbb{L}^2$ Riemannian metric.  
\end{proof}
Consequently, the geodesic distance on $\mathcal{H}$ is simply the $\mathbb{L}^2$ norm
\begin{align}
    d_{\mathcal{H}}(\boldsymbol{\theta}_0, \boldsymbol{\theta}_1) = \|\boldsymbol{\theta}_0 - \boldsymbol{\theta}_1\|_{\mathbb{L}^2}
\end{align}
and the geodesic path is the straight line connecting them
\begin{equation}
\alpha_\mathcal{H}(\tau) = (1-\tau) \boldsymbol{\theta}_0 + \tau \boldsymbol{\theta}_1.
\end{equation}

\begin{prop}
The map $\mathcal{R}_{\boldsymbol{\theta}} : F([0,1],\mathbb{R}^d) \to \Psi([0,1]) \times \mathcal{H}$, defined above, is a bijection.
\end{prop}
\begin{proof}
The element of $\Psi([0,1])$ uniquely defines the arc-length function by $s(t) = \int_0^t (\sqrt{\dot{s}(u)})^2du$. As mentioned before, we have a bijection between the unparametrized Frenet curvatures in $\mathcal{H}$ and the unparametrized curve in the shape space $\mathcal{S}$ (\ref{th:Frenet_ODE}, \ref{th:Fund_th_curves}). Then, from $X \in \mathcal{S}$, the initial parametrized curve is simply $x(t) = X(s(t))$.
\end{proof}
The set of such $\mathcal{R}_{\boldsymbol{\theta}}$ is the Cartesian product of  $\Psi([0,1])$ and $\mathcal{H}$ and, therefore, is also a Riemannian manifold equipped with the product metric $d_{\Psi} \oplus d_\mathcal{H}$ \cite{Math2005}. The induced metric on $F([0,1],\mathbb{R}^d)$ under the representation $\mathcal{R}_{\boldsymbol{\theta}}$ is 
\begin{align}
    d_{\boldsymbol{\theta}}(x_0,x_1) = d_{\Psi}\left(\sqrt{\dot{s}_0}, \sqrt{\dot{s}_1} \right) + d_{\mathcal{H}}(\boldsymbol{\theta}_0,\boldsymbol{\theta}_1) \,.
\end{align}
In order to define a distance on the shape space $\mathcal{S}$ from that one, we must quotient out the space $\textrm{Diff}_+([0,1])$. The action of $\textrm{Diff}_+([0,1])$ on $\Psi([0,1])$ is the same as \ref{eq:group_action_srvf}, and the Frenet curvatures are invariant under reparametrization of the corresponding parametrized curve (Lemma \ref{lem:invariance_curv_Q}). Moreover, by simply taking $h^* = s_1^{-1} \circ s_0 \in \textrm{Diff}_+([0,1])$, we have $\sqrt{\dot{s}_1} * h^* = \sqrt{\dot{s}_0}$, and thus the distance on $\Psi([0,1])/\textrm{Diff}_+([0,1])$ between $s_0$ and $s_1$ is zero. 
Hence, the induced distance on the shape space, between $X_0, X_1 \in \mathcal{S}$, under the representation $\mathcal{R}_{\boldsymbol{\theta}}$ is defined as 
\begin{equation}
d_{\mathcal{S}}^{(\boldsymbol{\theta})}(X_0,X_1) := d_{\mathcal{H}}(\boldsymbol{\theta}_0,\boldsymbol{\theta}_1) = \|\boldsymbol{\theta}_0 - \boldsymbol{\theta}_1\|_{\mathbb{L}^2}
\end{equation}
and the geodesic path connecting them is 
\begin{equation}
\alpha^{(\boldsymbol{\theta})}_\mathcal{S}(\tau) = \left( \sqrt{\dot{s}_0}, \, \alpha_\mathcal{H}(\tau) \right).
\end{equation}
This immediate representation by the Frenet curvatures appears in the experiments not to be sufficiently elastic (Figure~\ref{fig:geod_theta_boucle}). It has somewhat the same weakness as the angle representation proposed in \cite{BookFunctionalShapeDataAnalysis}, the Frenet curvatures being already independent of the parametrization. 


\subsection{Square Root Curvatures Transform}

To overcome the \say{non-elasticity} issue of the representation defined above, we propose a second framework for shape analysis based on Frenet curvatures which uses, like the square root velocity function, the parametrization as a tool to register the curves and define a more \say{elastic} method. The latter is inspired by the square-root velocity transform of $SO(d)$-valued curves.

\begin{definition}[SRV Transform for curves on $SO(d)$]
Let $P \in C^{\infty}([0,1],SO(d))$. The Square Root Velocity transform of $P$ is the map
\begin{equation}
q(P)(t) = \frac{L_{P(t)^{-1}}\dot{P}(t)}{\sqrt{\|\dot{P}(t)\|_F}} =  \frac{P(t)^T\dot{P}(t)}{\sqrt{\|\dot{P}(t)\|_F}}\,,
\end{equation}
where $\|.\|_F$ is the Frobenius norm associated with the scalar product on the Lie Algebra of skew-symmetric matrices $\langle A, B \rangle = \frac{1}{2}tr(A^T B) = -\frac{1}{2}tr(AB)$.
\end{definition}
Let $x\in F([0,1], \mathbb{R}^d)$ and $Q(t) \in C^{\infty}([0,1],SO(d))$ be its associated Frenet path. Using the Frenet-Serret differential equation, the SRV Transform of the Frenet path is 
\begin{equation}
q(Q)(t) = \sqrt{\dot{s}(t)}\frac{A_{\boldsymbol{\theta}}(s(t))}{\sqrt{\|A_{\boldsymbol{\theta}}(s(t))\|_F}} \,.
\end{equation}
\begin{prop} \label{prop:Atheta}
Let $\boldsymbol{\theta} \in \mathcal{H}$, we have 
\[\|A_{\boldsymbol{\theta}}(s(t))\|_F = \|\boldsymbol{\theta}(s(t))\|_{2} \,. \] 
\end{prop}
Based on the SRV Transform of a Frenet path and Proposition~\ref{prop:Atheta}, we propose a new transformation of a parametrized curve, which we have called the \textit{Square-Root Curvatures (SRC)} transform. 
\begin{definition}[Square-Root Curvatures Transform]
Let $x\in F([0,1],\mathbb{R}^d)$. We consider its associated arc-length function $s(t)$ and Frenet curvatures $\boldsymbol{\theta}(s(t))$ defined as in Theorem~\ref{th:Frenet_ODE}. Then we define its square-root curvatures transform to be the map 
\begin{equation}
c(t) = \sqrt{\dot{s}(t)}\frac{\boldsymbol{\theta}(s(t))}{\sqrt{\|\boldsymbol{\theta}(s(t))\|}} \,.
\end{equation}
\end{definition}
The set of such square-root curvatures transforms is 
\begin{equation}
\mathcal{C} = \left \{ c \in \mathbb{L}([0,1],\mathbb{R}^{d-1}) \, | \, c_1, \hdots, c_{d-2} > 0  \right \} \,,
\end{equation}
which is the same as the set of admissible Frenet curvatures $\mathcal{H}$. We have already shown in the previous section that this set is a Riemannian manifold equipped with the $\mathbb{L}^2$ metric. Therefore, the geodesic distance between $c_0, c_1 \in \mathcal{C}$ is the $\mathbb{L}^2$ distance between them, and the geodesic path is a straight line. 
We define the following representation of a parametrized curve $x\in F([0,1],\mathbb{R}^d)$, from its Square-Root Curvatures transform, by
\begin{equation}
\mathcal{R}_{\textrm{SRC}}(x)(t) = \left( \sqrt{\dot{s}(t)}, \, c(t) \right) \,.
\end{equation}
\begin{prop}
The map $\mathcal{R}_{\textrm{SRC}} : F([0,1],\mathbb{R}^d) \to \Psi([0,1]) \times \mathcal{C}$, defined above, is a bijection.
\end{prop}
\begin{proof}
This is again a result of theorems \ref{th:Frenet_ODE} and \ref{th:Fund_th_curves}. To get $x$ from $\mathcal{R}_{\textrm{SRC}}(x)$, it should be noted firstly that $c(t)\|c(t)\| = \dot{s}(t)\boldsymbol{\theta}(s(t))$. From that, the skew-symmetric matrix function of the Frenet-Serret ODE can be reconstructed. By solving the corresponding Frenet-Serret ODE one gets the associated time parametrized Frenet path $Q(t)$. Then, using the first component of $\mathcal{R}_{\textrm{SRC}}(x)$, we get $x(t) = X(s(t)) = \int_0^t \dot{s}(u)T(s(u))du$.  
\end{proof}
The set of such square root curvature representations $\mathcal{R}_{\textrm{SRC}}$ is the Cartesian product $\Psi([0,1]) \times \mathcal{C}$ and therefore a Riemannian manifold with the product metric $d_{\Psi} \oplus d_\mathcal{C}$. This representation is, by definition, invariant under the action of $SO(d)$. Then, the corresponding shape space is the quotient space $\Psi([0,1]) \times \mathcal{C} / \textrm{Diff}([0,1])$. Let's $x \in F([0,1],\mathbb{R}^d)$ and $h \in \textrm{Diff}([0,1])$. The SRC representation of $\Tilde{x} = x \circ h$ is 
\begin{equation}
\mathcal{R}_{\textrm{SRC}}(\Tilde{x}) = \left(\sqrt{\dot{s}} * h, \, c * h  \right) = \mathcal{R}_{\textrm{SRC}}(x) * h 
\end{equation}
where $*$ is the group action defined in \ref{eq:group_action_srvf}. 
\begin{prop}
The metric on $F([0,1],\mathbb{R}^d)$ induced by the Riemannian metric on $\Psi([0,1]) \times \mathcal{C}$ defined by $d_{\textrm{SRC}} := d_{\Psi} \oplus d_\mathcal{C}$ is invariant under the action of $\textrm{Diff}_+([0, 1])$.
\end{prop}
The distance on the shape space $\mathcal{S}$ under the representation $\mathcal{R}_{\textrm{SRC}}$, between two elements $X_0,X_1 \in \mathcal{S}$, is defined as
\begin{equation}
d_{\mathcal{S}}^{(\textrm{SRC})}(X_0, X_1) := \inf_{h\in \textrm{Diff}_+([0,1])} d_{\textrm{SRC}}(x_0, x_1 \circ h).
\end{equation}
\noindent From the optimal wrapping function $h^*$ the geodesic path on $\mathcal{S}$ between them is 
\begin{equation}
\begin{array}{rc}
     \alpha^{(\textrm{SRC})}_\mathcal{S}(\tau) = &\left( \frac{\sin((1-\tau)\vartheta)}{\sin(\vartheta)} \sqrt{\dot{s}_0} +  \frac{\sin(\tau\vartheta)}{\sin(\vartheta)}(\sqrt{\dot{s}_1} * h^*), \right. \\
     & \left. (1-\tau)c_0 + \tau (c_1 * h^*) \right) \,
\end{array}
\end{equation}
\noindent where $\vartheta = d_{\Psi}(\sqrt{\dot{s}_0}, \sqrt{\dot{s}_1} * h^* )$.The registration problem consider here is to find the minimizer $h^*$ over $\textrm{Diff}_+([0,1])$ of 
\begin{align}
\int_0^1 \|c_0(t) - (c_1 * h)(t) \|^2 + \| \sqrt{\dot{s}_0(t)} - (\sqrt{\dot{s}_1} * h)(t) \|^2  dt \,.
\end{align}
Using the reformulation principle of \cite{Brunel2019}, that is $\gamma = s_1 \circ h \circ s_0^{-1} \in \textrm{Diff}_+([0,1])$, this registration problem is shown to be equivalent to finding $\gamma^* \in \textrm{Diff}_+([0,1])$ that minimizes
\begin{multline}
    \int_0^1 \left \|\frac{\boldsymbol{\theta}_0(s)}{\sqrt{\|\boldsymbol{\theta}_0(s)\|}} - \sqrt{\gamma'(s)}\frac{\boldsymbol{\theta}_1(\gamma(s))}{\sqrt{\|\boldsymbol{\theta}_1(\gamma(s))\|}} \right\|^2 \\ + \|1 - \gamma'(s)\|^2ds \,.
\end{multline}
Note that this reformulation has the form of a penalized registration problem. The second term represents a penalty term on $\gamma$ and ensures a certain smoothness of the warping function. In this framework, the deformable templates are the square-root normalized curvatures which encode more geometric information than the unit tangent vector.


\section{Experiments}

In this section, we report the experimental results of the proposed methods, comparing them with the SRVF method. We use both synthetic and real data. Additional results and figures are available in the supplementary material. 

\subsection{Statistical estimation of the Frenet curvatures}

The main limitation of shape analysis methods based on the Frenet curvatures is the need of additional estimates of curvatures. Being dependent on higher-order derivatives (up to order $d$), they are quite sensitive to the observation noise of the Euclidean curve. We detail here a simple method that can be used for their smooth estimation, and we refer to \cite{Sangalli2009, Park2022} for a more complex and detailed statistical estimation algorithm. First, it is possible to use a local polynomial smoothing algorithm to estimate the $d$ first derivatives of the Euclidean curve \cite{Fan1996}. From these derivatives, the raw estimates of the Frenet curvatures can be computed by using their extrinsic formulas. 
We propose here a second method based on the Frenet-Serret ODE approximation, to obtain the raw estimates, that appear to be more stable. A simple middle point approximation of the ODE solution gives
\begin{equation}
Q(s_j) \approx Q(s_i)\exp\left((s_i - s_j)A_{\boldsymbol{\theta}}\left(\frac{s_i + s_j}{2}\right) \right).
\end{equation}
where $\exp(.)$ is the exponential map of the Lie group $SO(d)$. Then, using the inverse logarithm map $\log(.)$, this gives an approximation of the matrix $A_{\boldsymbol{\theta}}\left((s_i + s_j)/2\right) \approx \frac{1}{s_i-s_j}\log\left( Q(s_i)^T Q(s_j)\right)$ and by identification raw estimates of $\boldsymbol{\theta}\left((s_i + s_j)/2\right)$. As we consider here a problem of estimating a functional parameter, we formulate the final $\boldsymbol{\theta}$ estimation problem as a penalized weighted functional regression with the obtained raw estimates, that we solve by using a B-spline approximation of $\boldsymbol{\theta}$. 

\subsection{Experiments with synthetic curves}

We use synthetic data to highlight the differences between the methods discussed above (SRVF, SRC, and Frenet curvatures). 
The computations related to the SRVF method are made with the package \href{https://fdasrsf-python.readthedocs.io/en/latest/}{fdasrsf}. 
The SRC and Frenet curvatures methods are implemented with the code provided as supplementary material, including a dynamic programming algorithm for solving the registration problems.

We consider the simple case of a set of $20$ curves in $\mathbb{R}^2$ with a single large peak of curvature. This one is created by generating curvature functions on $[0,1]$ with one peak of maximum value $60.5$, width $0.15$, and location chosen randomly between $0.1$ and $0.9$. These curves have the shape of a loop made with a wire, where the loop is more or less close to the right or left wire end, depending on the location of the curvature peak. 
\begin{figure}[h]
\begin{center}
\includegraphics[width=1\linewidth]{ 
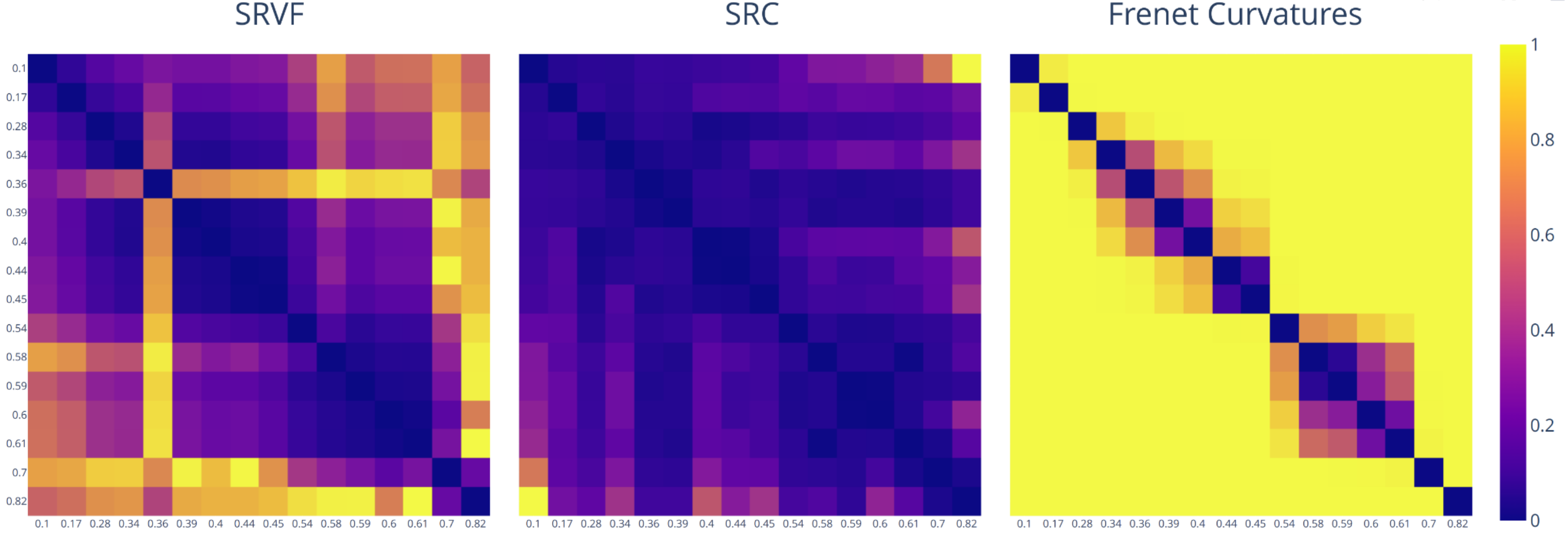}
\end{center}
   \caption{Matrices of pairwise SRVF distance (left), SRC distance (middle), and unparametrized Frenet curvatures distance (right) by sorted location of the peak curvature.}
\label{fig:heatmaps_2D}
\end{figure}
\begin{figure}[h]
\begin{center}
\includegraphics[width=1\linewidth]{ 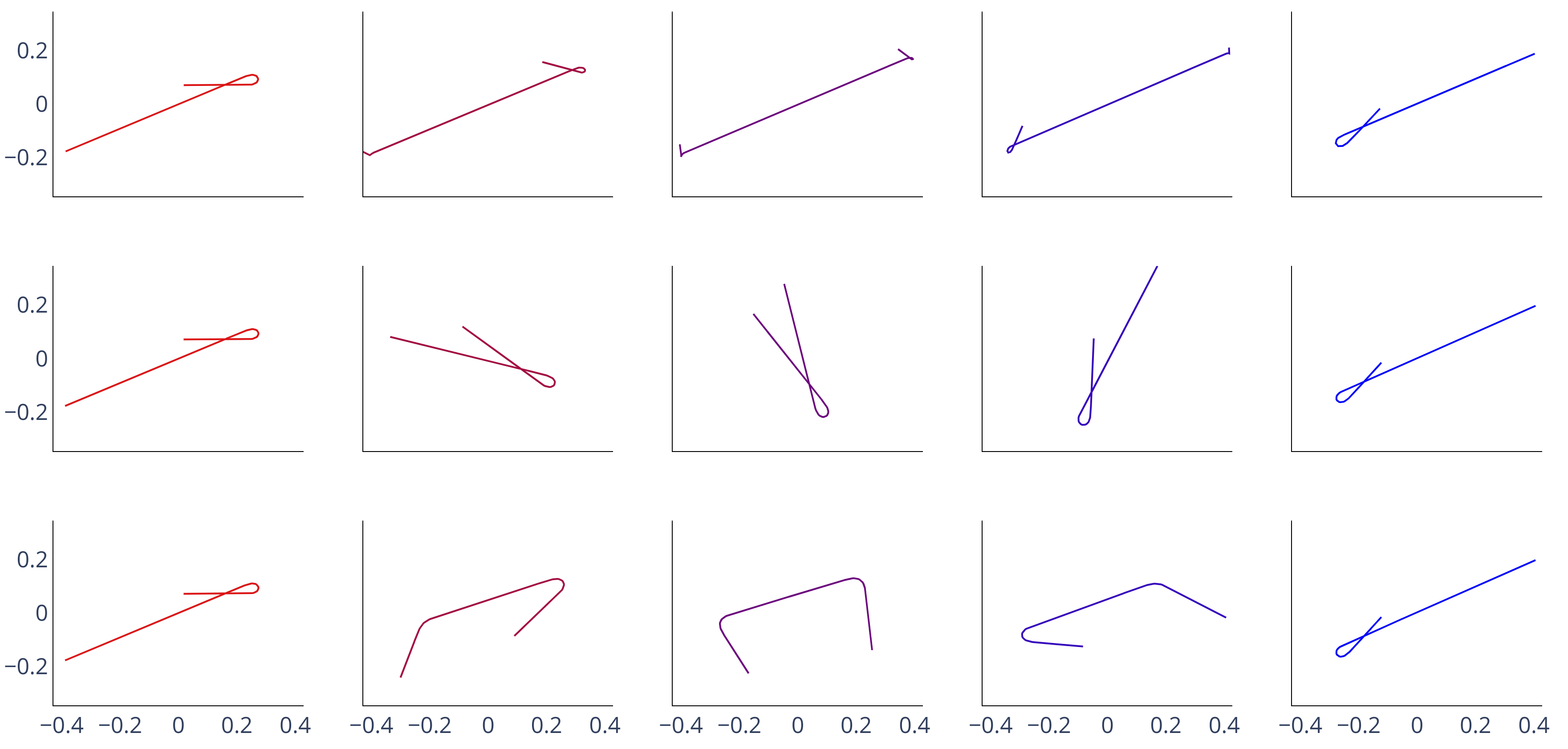}
\end{center}
   \caption{SRVF ($1^{st}$ row), SRC ($2^{nd}$ row), and Frenet curvatures ($3^{rd}$ row) geodesic paths between curves with curvature peaks located at $0.27$ (left) and $0.78$ (right).}
\label{fig:geod_x_boucle}
\end{figure}
\begin{figure}[h]
\begin{center}
\includegraphics[width=1\linewidth]{ 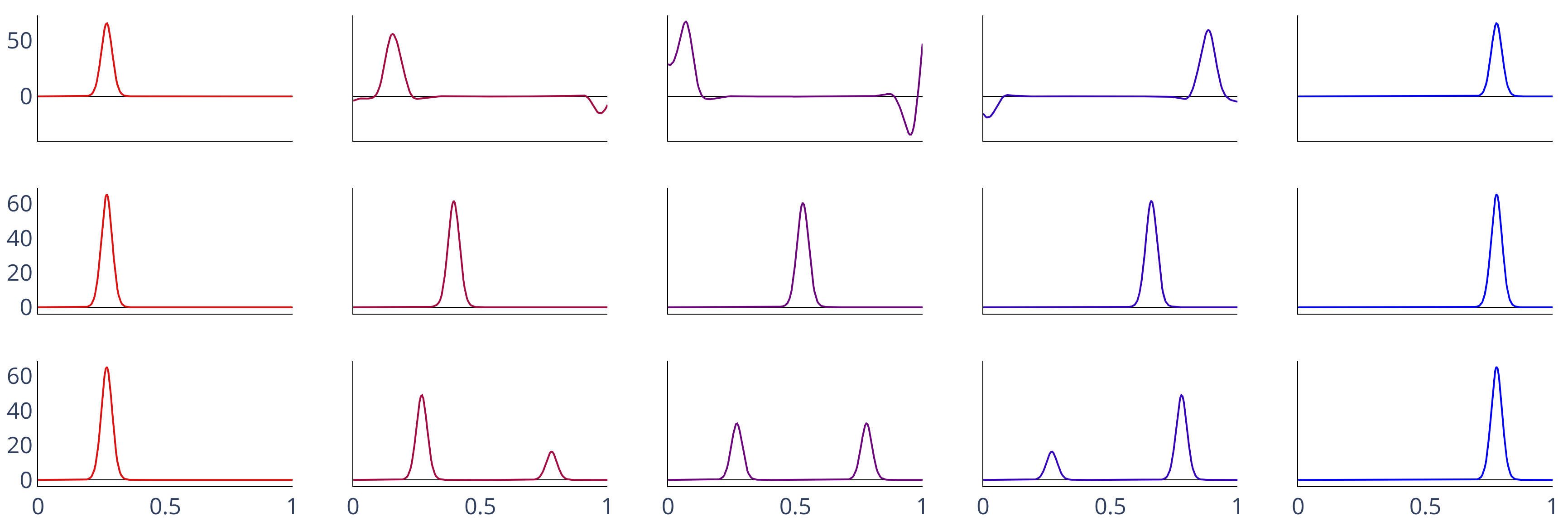}
\end{center}
   \caption{Curvatures of the Euclidean curves along the geodesic paths plotted in Figure \ref{fig:geod_x_boucle}: SRVF ($1^{st}$ row), SRC ($2^{nd}$ row), and Frenet curvatures ($3^{rd}$ row).}
\label{fig:geod_theta_boucle}
\end{figure}

We compare the three methods through the pairwise distance matrices in Figure~\ref{fig:heatmaps_2D}, and the geodesic paths computed between two of these curves in Figure~\ref{fig:geod_x_boucle}, with peaks located at $0.27$ (red curve) and at $0.78$ (blue curve). 
The corresponding deformations through the variations of the curvature along the different geodesic paths on Euclidean curves are shown in Figure~\ref{fig:geod_theta_boucle}, which highlights the strengths and weaknesses of each method. First, it emphasizes the "non-elasticity" of the unparametrized Frenet curvatures method, as in the middle of the geodesic path, we have two peaks of curvature and, therefore, a completely different shape without any loop. This explains the inconsistency of the heatmap under this method. Conversely, there is an elastic deformation of the curvature with the SRC transform, and shapes along the geodesic are consistent with the set of curves considered, which is well summarised on the corresponding heatmap where all distances are rather close to zero. For the SRVF method, the chosen example with peaks of curvature that are quite far apart shows that artifacts appear along the geodesic; the middle curve has two small loops at the edges. This phenomenon gives unreliable distances, as shown in the heatmap, where the distances are not monotone as a function of the spacing between the curvature peaks. 


By considering a set of curves characterized by specific features, we observe a clear difference in consistency of the shapes along the geodesics with respect to the different methods. This phenomenon is quite visible on the geodesic between helices in 2D, Figure~\ref{fig:geod_spiral} or 3D, Figure~\ref{fig:geod_x_helix}.
In that case, within both geodesic paths under SRVF method, the curves lose the characteristic geometry of the helix. A three-dimensional circular helix is characterized by having a constant curvature and torsion, which is not the case along the SRVF geodesic, but preserved with the SRC method. 
In that case the geodesic under the Frenet curvature representation is very similar to SRC. 




\subsection{Application to sign language motion data}

\begin{figure}[h]
\begin{center}
\begin{subfigure}[b]{0.48\linewidth}
    \includegraphics[width=1\linewidth]{ 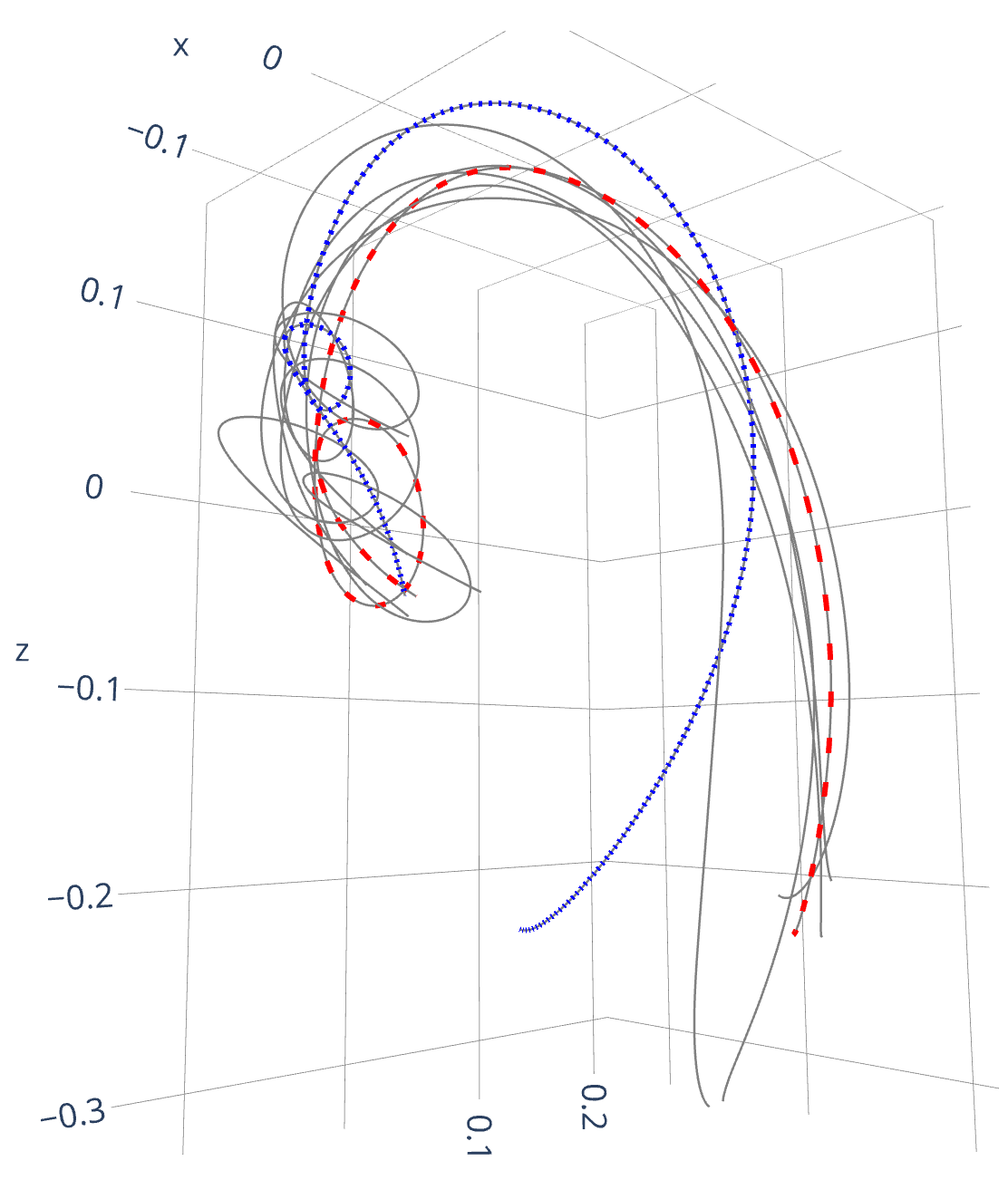}
    \end{subfigure} 
    \begin{subfigure}[b]{0.5\linewidth}
    \includegraphics[width=1\linewidth]{ 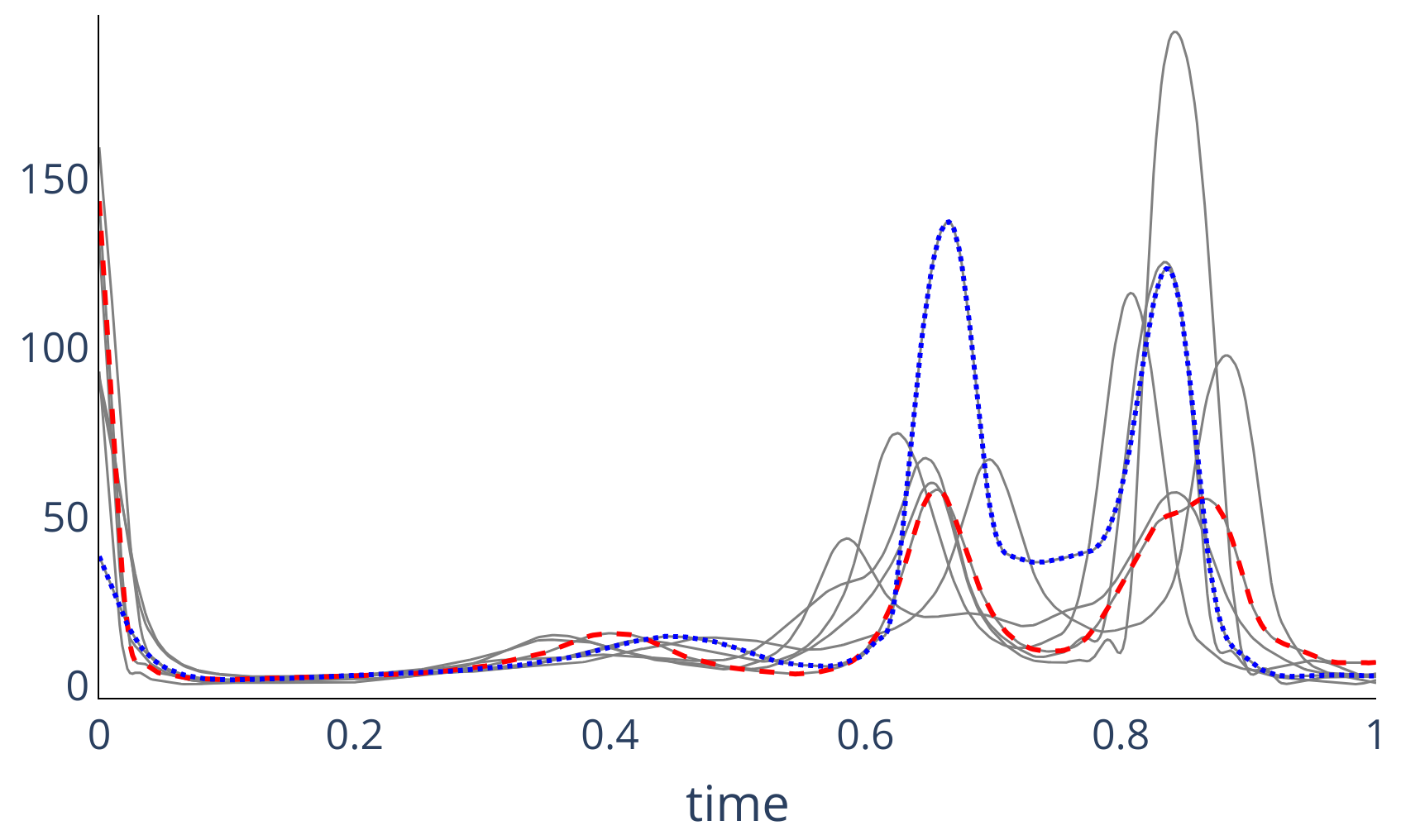} 
    \includegraphics[width=1\linewidth]{ 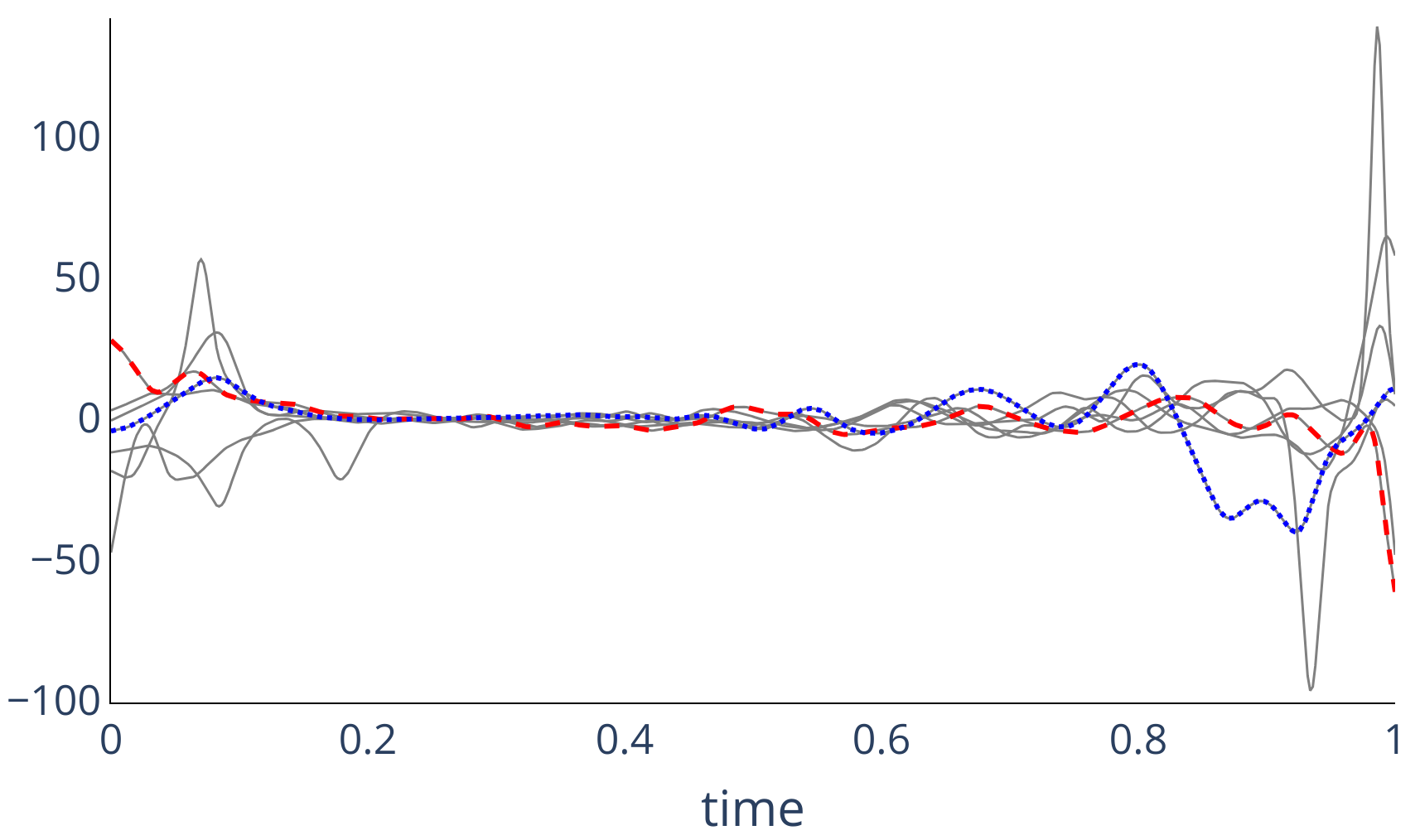}
\end{subfigure}
\end{center}
   \caption{Trajectories of the right wrist while signing "Femme" in sign language: 3D curves (left), time-parametrized curvatures (top right), torsions (bottom right). The blue and red ones are used to compute the geodesic in Figure~\ref{fig:goed_femme}.}
\label{fig:init_femme}
\end{figure}

It appears that curvilinear velocity and Frenet curvatures are particularly relevant parameters for the analysis of human motion. Several laws involving these parameters can be found in the literature \cite{Lacquaniti1983, Maoz2009, Pollick2009, FlashBerthoz2021}; among others, the power laws state a special relationship between the curvature, the torsion, and the velocity of a point trajectory representing human motion. Using a method that conserves the shape of these parameters is, therefore, of particular interest in this application. 
We demonstrate here with the case of wrist trajectories in sign language, acquired with a motion capture system by the company MocapLab (\href{https://www.mocaplab.com/fr/}{https://www.mocaplab.com/fr/}). We compute the geodesic paths, under each of the frameworks, between the arbitrarily chosen red and blue curves within the set of several repetitions of the sign "Femme", shown in Figure~\ref{fig:init_femme} with the corresponding time-parameterized Frenet curvatures. 
\begin{figure*}
\begin{center}
\begin{subfigure}[c]{0.3\linewidth}
\begin{center}
    \includegraphics[width=1\linewidth]{ 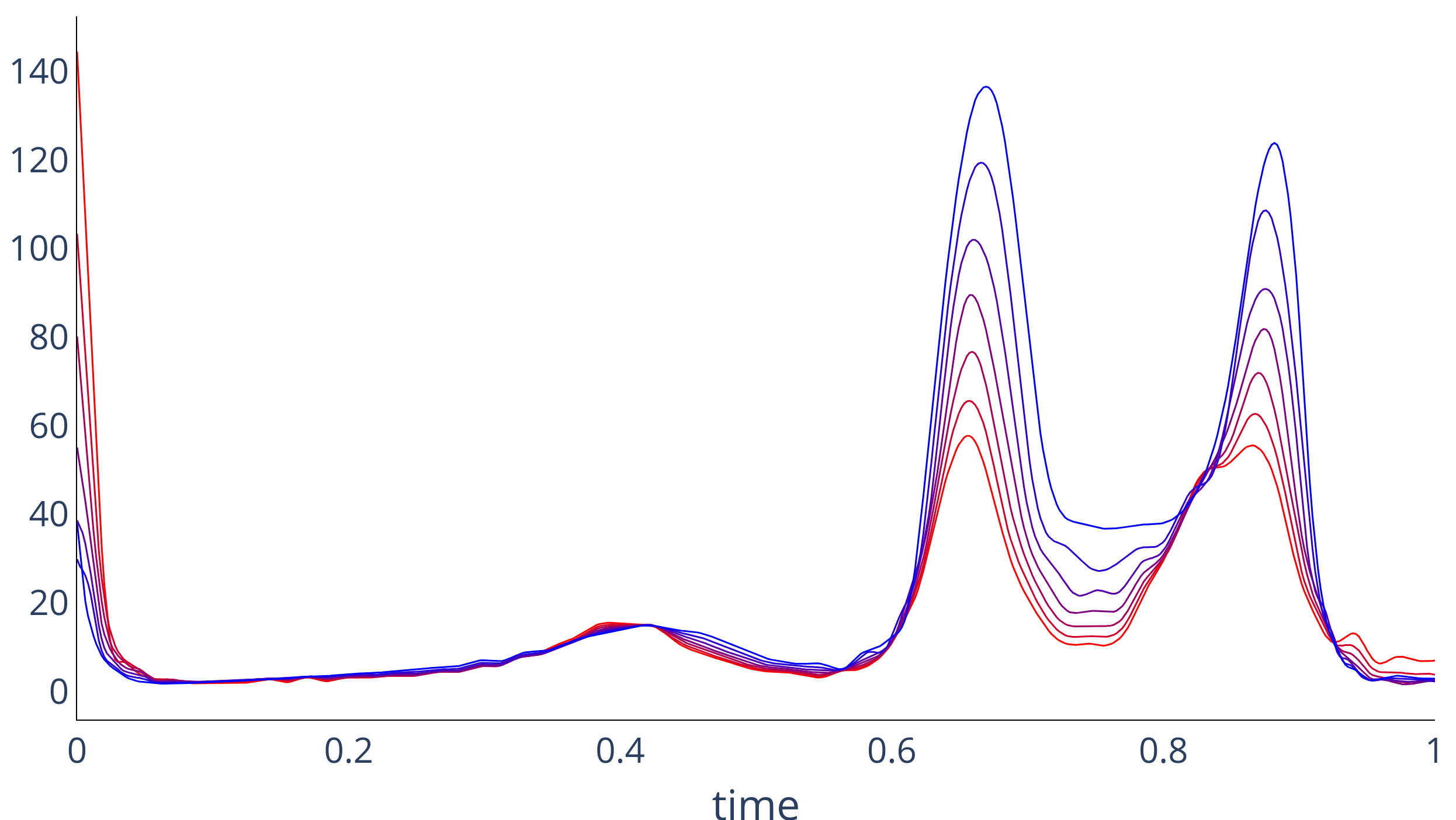} \\
    \includegraphics[width=1\linewidth]{ 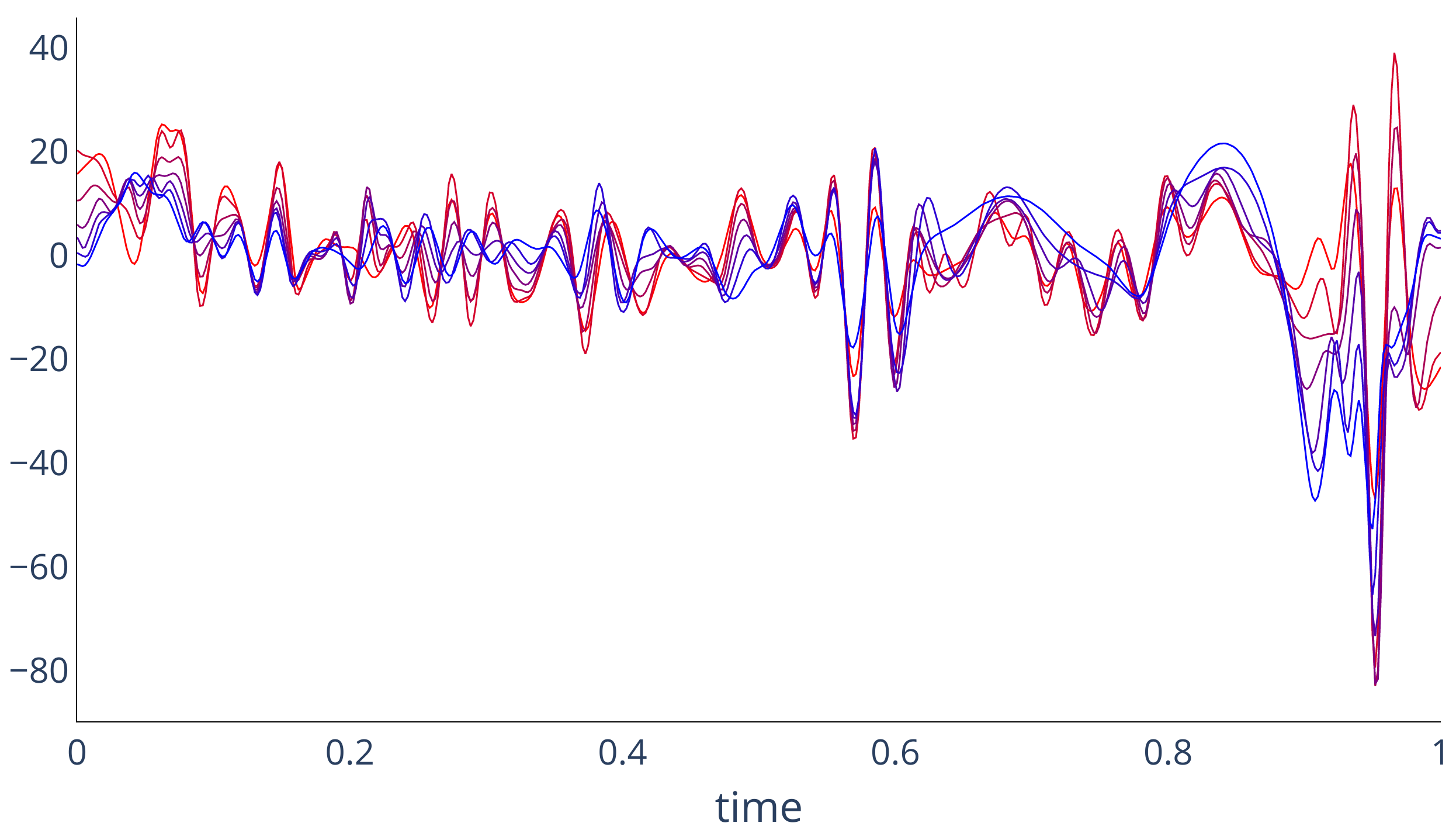}
\end{center}
\subcaption{SRVF}
\end{subfigure} 
\hspace{0.2cm}
\begin{subfigure}[c]{0.3\linewidth}
\begin{center}
    \includegraphics[width=1\linewidth]{ 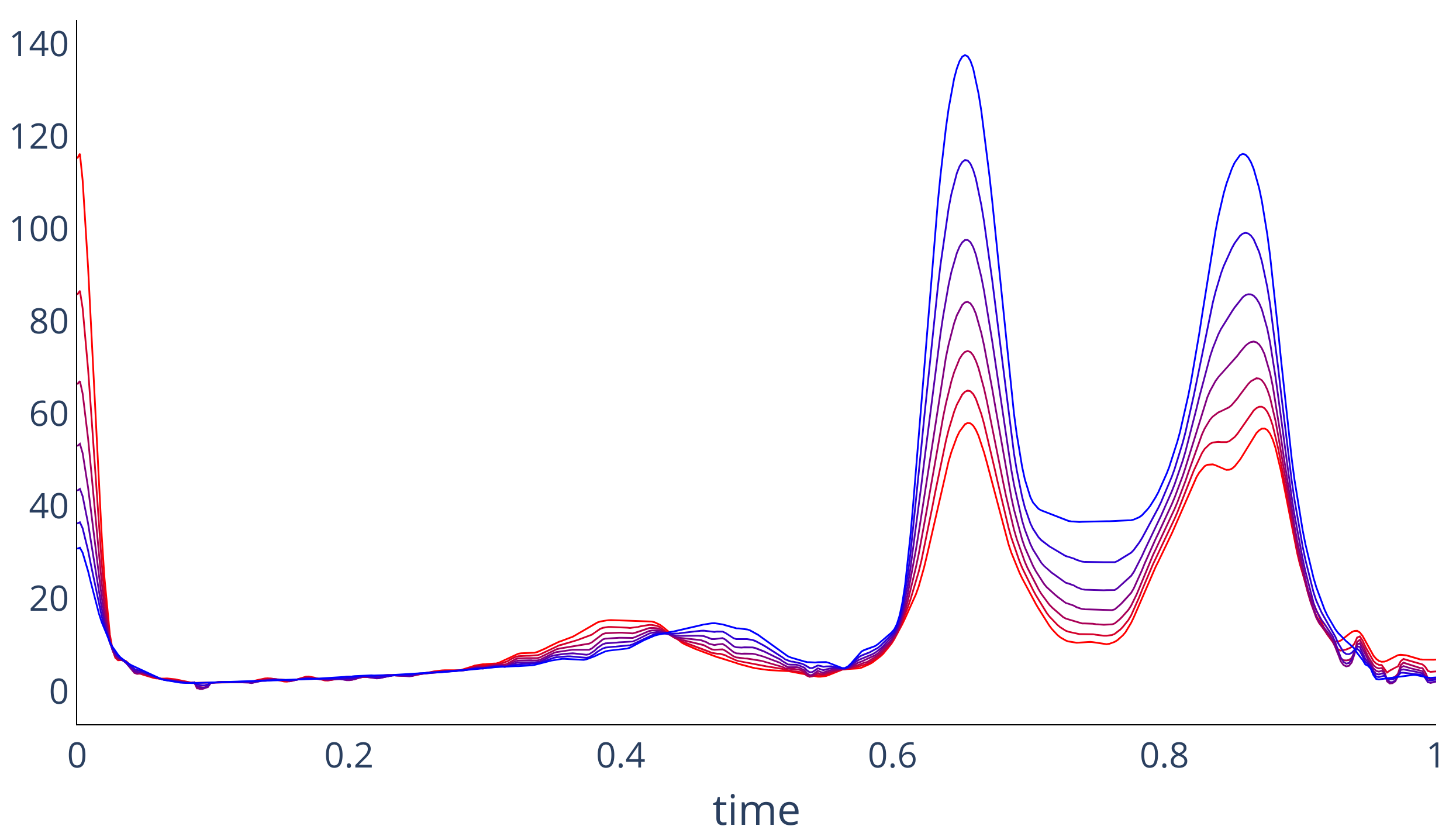} \\
    \includegraphics[width=1\linewidth]{ 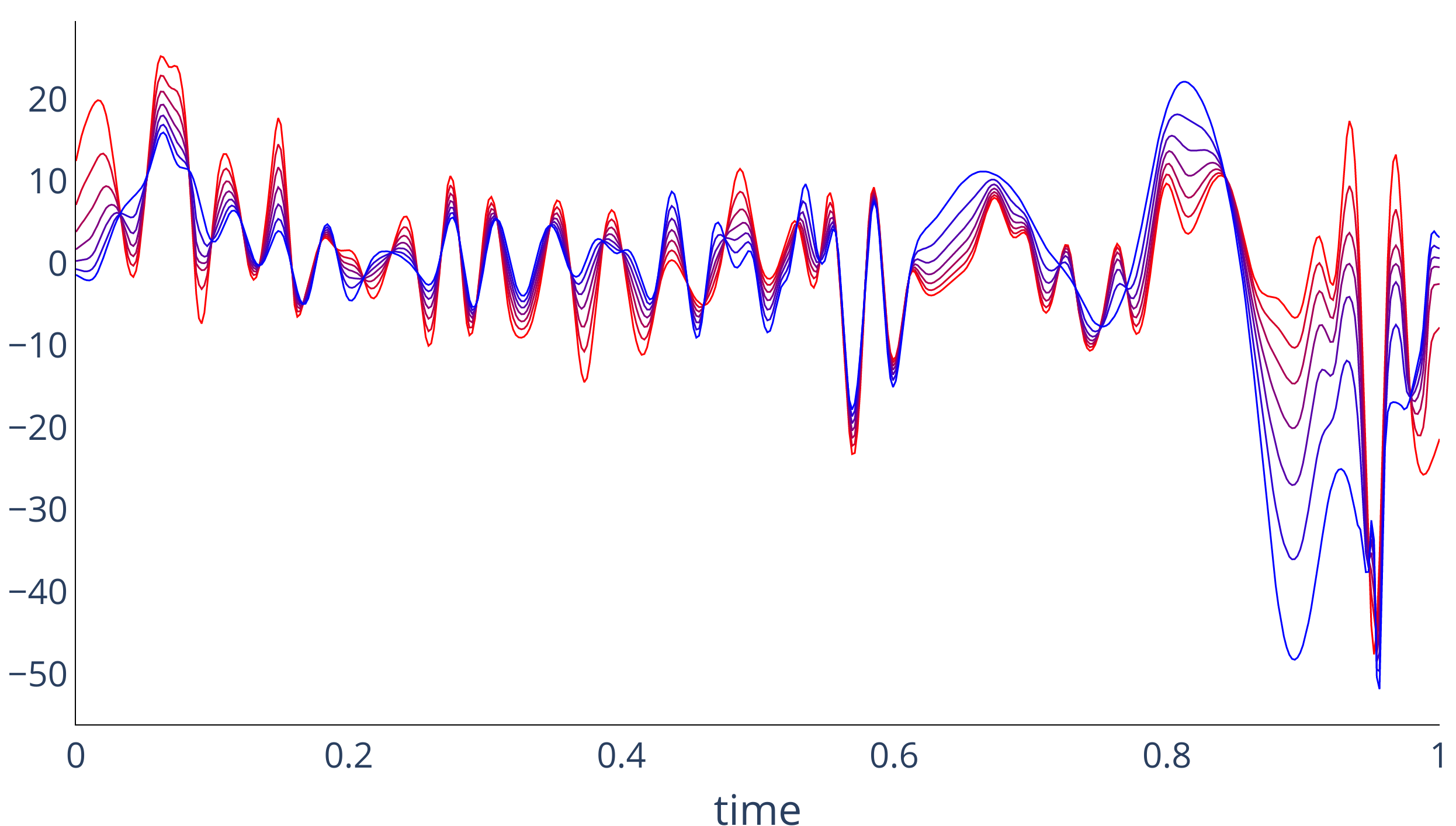}
\end{center}
\subcaption{SRC}
\end{subfigure} 
\hspace{0.2cm}
\begin{subfigure}[c]{0.3\linewidth}
\begin{center}
    \includegraphics[width=1\linewidth]{ 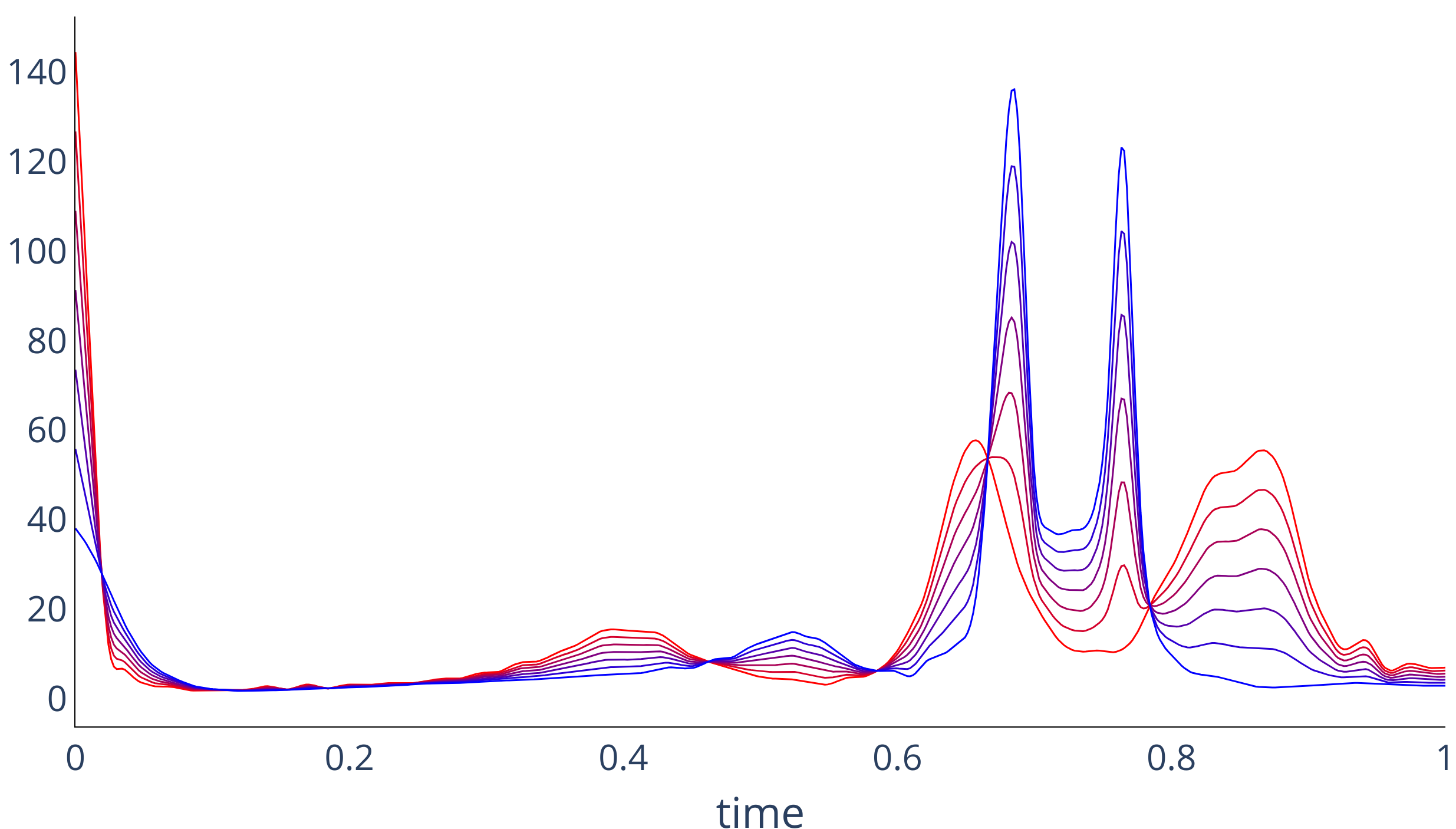} \\
    \includegraphics[width=1\linewidth]{ 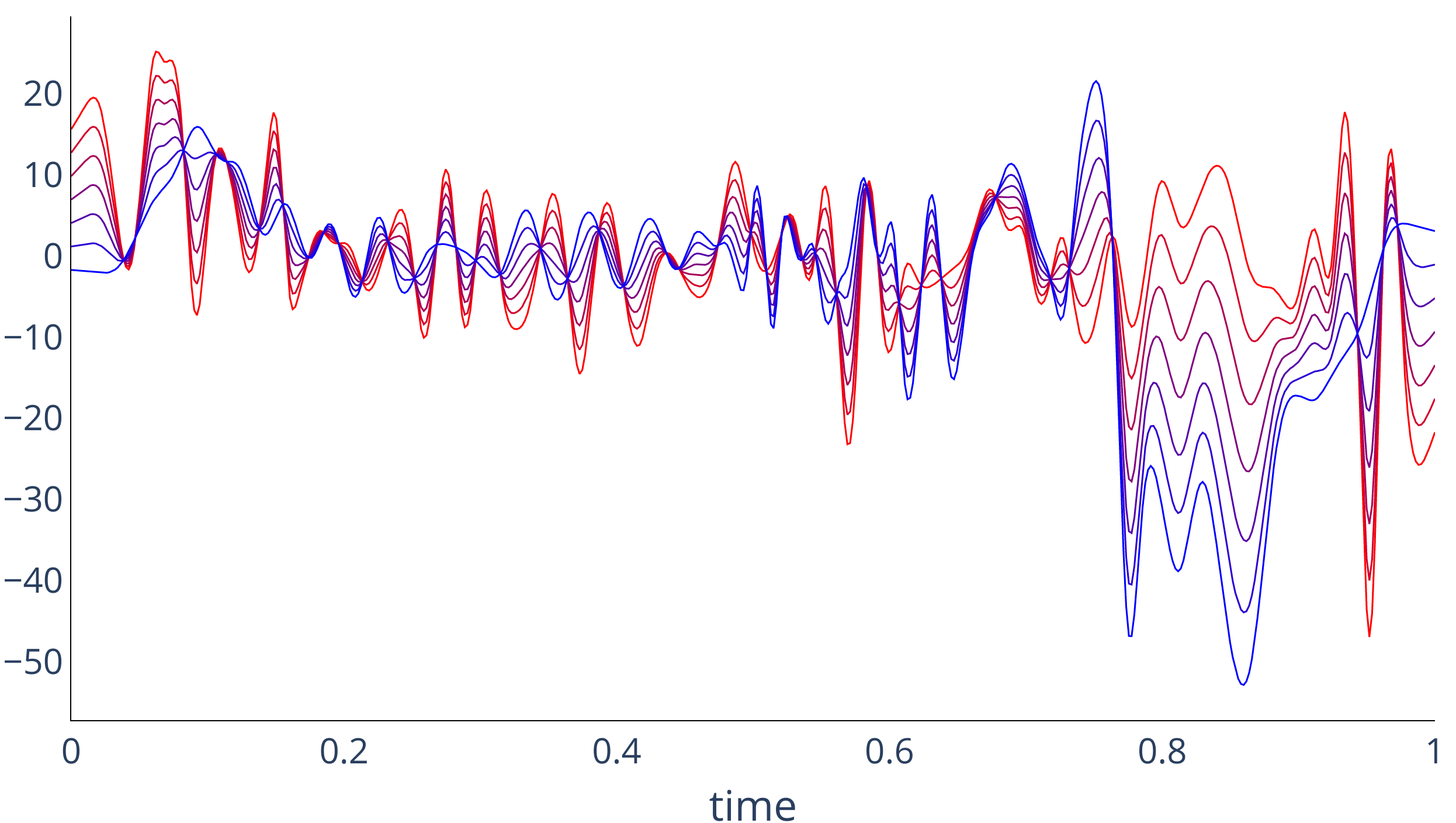}
\end{center}
\subcaption{Frenet curvatures}
\end{subfigure} 
\end{center}
\vspace{-0.3cm}
   \caption{Comparison between time-parametrized curvature and torsion along the geodesic path under SRVF (left), SRC (middle), and Frenet curvatures (right).}
\label{fig:goed_femme}
\end{figure*}
\begin{figure}[h]
\centering
\begin{subfigure}[c]{0.45\linewidth}
\begin{center}
    \includegraphics[width=0.9\linewidth]{ 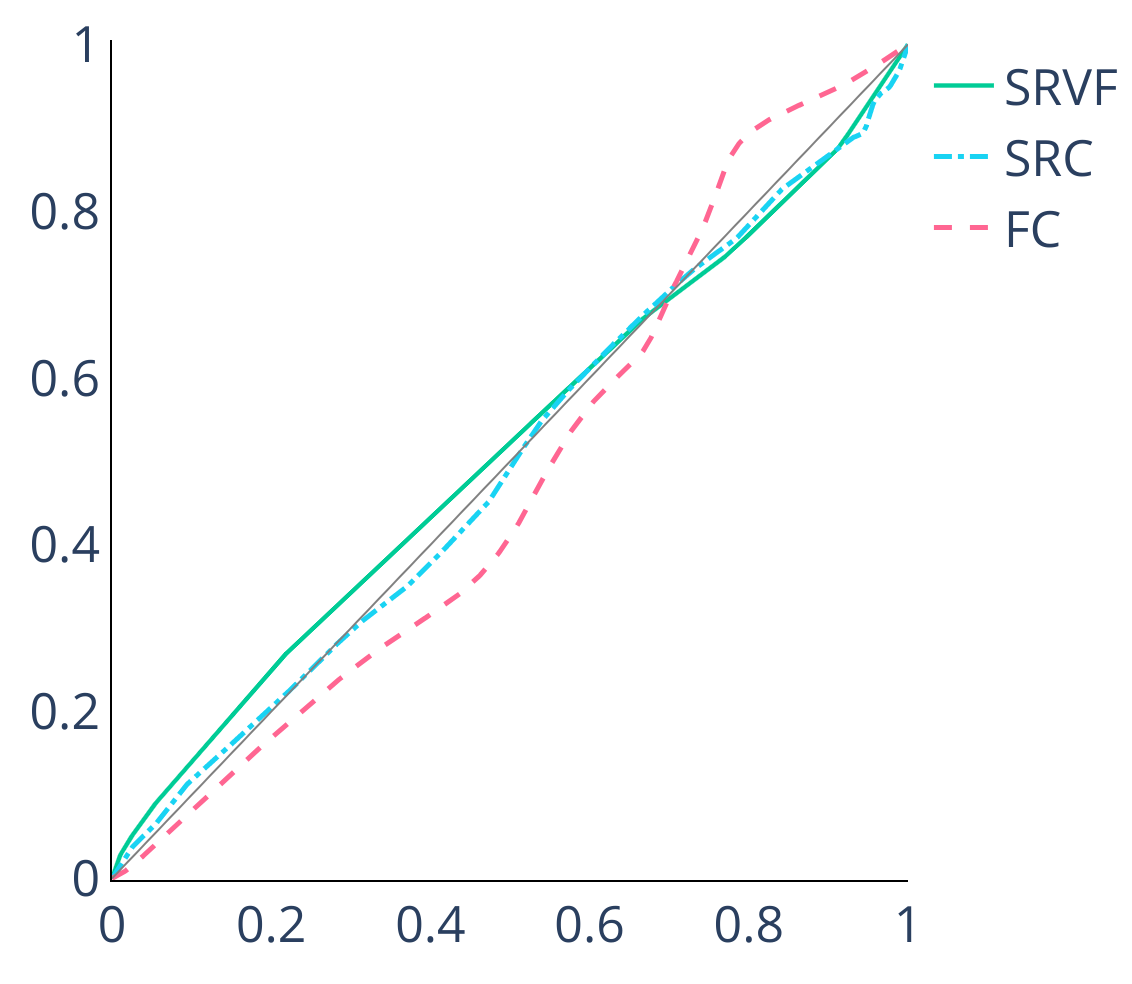}
\end{center}
\subcaption{Warping functions $h$}
\end{subfigure} 
\begin{subfigure}[c]{0.45\linewidth}
\begin{center}
    \includegraphics[width=0.9\linewidth]{ 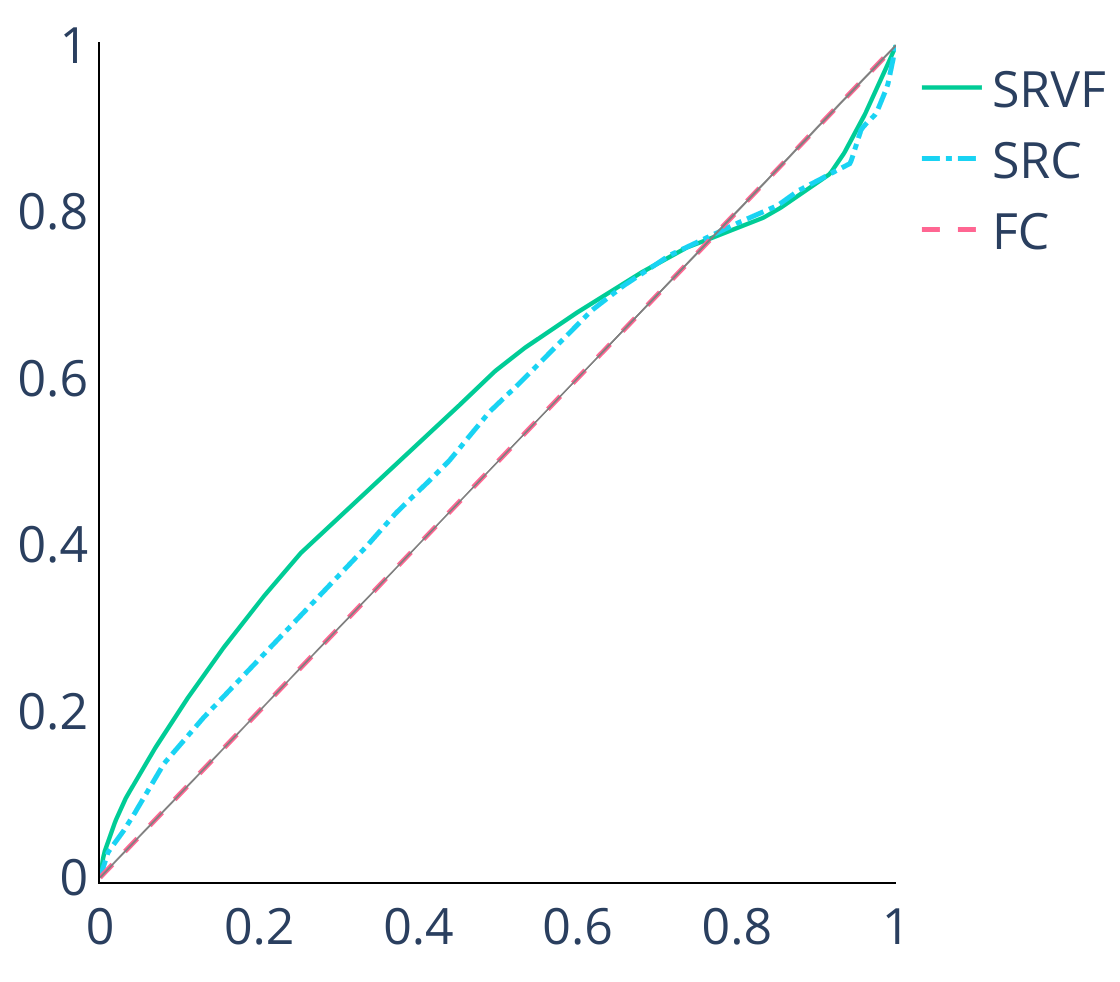}
\end{center}
\subcaption{Warping functions $\gamma$}
\end{subfigure} 
   \caption{Comparison of estimated warping functions $h$ (left) and $\gamma$ (right) to compute the geodesic in Figure~\ref{fig:goed_femme}.}
\label{fig:warp_fct_femme}
\end{figure}

Figure~\ref{fig:goed_femme} emphasizes the advantage of considering a representation depending on the parameterization,
allowing a registration before computing the geodesic. However, it also shows that considering only the tangent vector as a representative object (SRVF) is not sufficient to find the optimal reparametrization that correctly aligns the torsions, and could affect subsequent analyses made by the SRVF method, such as the mean. This results in the appearance of new minimums, maximums and zeros in the torsion functions along the SRVF geodesic. However, such characteristic points of the curvature, torsion and velocity functions are crucial in the observation of the laws of motion. It is therefore preferable to use a method that preserves these characteristics by directly optimizing the optimal alignment from these parameters, such as the proposed SRC method.  


\section{Conclusion}

The square-root curvature transform of a Euclidean curve in $\mathbb{R}^d$ is a representation that encodes more geometric information of the curves, and thus the results are easier to interpret than existing methods. 
The main limitation lies in the estimation of the Frenet curvatures from real and noisy data, nevertheless recent smooth statistical estimators can be used for computing the SRC \cite{Sangalli2009, Park2022}. 
We believe our method is particularly interesting for motion trajectory analysis and could be developed further in the future as a tool for generation, segmentation and classification of complex trajectories.

{\small
\bibliographystyle{ieee_fullname}
\bibliography{egbib}

@book{BookFunctionalShapeDataAnalysis,
  title={Functional and Shape Data Analysis},
  author={Srivastava, Anuj and Klassen, Eric},
  isbn={9781493981557},
  series={Springer Series in Statistics},
  url={https://books.google.fr/books?id=Y6GaugEACAAJ},
  year={2016},
  publisher={Springer New York}
}

@article{Srivastava2011,
  author={Srivastava, Anuj and Klassen, Eric and Joshi, Shantanu H. and Jermyn, Ian H.},
  title={Shape Analysis of Elastic Curves in Euclidean Spaces}, 
  journal={IEEE Transactions on Pattern Analysis and Machine Intelligence}, 
  year={2011},
  volume={33},
  number={7},
  pages={1415-1428},
  doi={10.1109/TPAMI.2010.184}}

@InProceedings{Brunel2019,
author="Brunel, Nicolas J.-B.
and Park, Juhyun",
editor="Nielsen, Frank
and Barbaresco, Fr{\'e}d{\'e}ric",
title="The Frenet-Serret Framework for Aligning Geometric Curves",
booktitle="Geometric Science of Information",
year="2019",
publisher="Springer International Publishing",
address="Cham",
pages="608--617",
isbn="978-3-030-26980-7"
}

@article{Bauer2021,
   author = {Martin Bauer and Nicolas Charon and Eric Klassen and Alice Le Brigant},
   doi = {10.1007/978-3-030-03009-4_87-1},
   journal = {Handbook of Mathematical Models and Algorithms in Computer Vision and Imaging},
   pages = {1-35},
   title = {Intrinsic Riemannian Metrics on Spaces of Curves: Theory and Computation},
   year = {2021},
}

@misc{Bauer2022,
	title = {Elastic {Metrics} on {Spaces} of {Euclidean} {Curves}: {Theory} and {Algorithms}},
	shorttitle = {Elastic {Metrics} on {Spaces} of {Euclidean} {Curves}},
	url = {http://arxiv.org/abs/2209.09862},
	doi = {10.48550/arXiv.2209.09862},
	urldate = {2023-03-04},
	publisher = {arXiv},
	author = {Bauer, Martin and Charon, Nicolas and Klassen, Eric and Kurtek, Sebastian and Needham, Tom and Pierron, Thomas},
	year = {2022},
	note = {arXiv:2209.09862 [math]},
}

@misc{Younes2018,
  title={Elastic distance between curves under the metamorphosis viewpoint},
  author={Laurent Younes},
  note={arXiv:1804.10155},
  year={2018}
}

@book{Younes2000,
   author = {Laurent Younes},
   isbn = {9783642120541},
   publisher = {Springer},
   series = {Applied Mathematical Sciences},
   pages = {441},
   title = {Shapes and Diffeomorphisms},
   year = {2010},
}

@article{Younes1998,
	title = {Computable {Elastic} {Distances} between {Shapes}},
	volume = {58},
	issn = {0036-1399},
	number = {2},
	urldate = {2023-03-07},
	journal = {SIAM Journal on Applied Mathematics},
	author = {Younes, Laurent},
	year = {1998},
	note = {Publisher: Society for Industrial and Applied Mathematics},
	pages = {565--586},
}

@article{Marron2015,
   author = {J. S. Marron and James O. Ramsay and Laura M. Sangalli and Anuj Srivastava},
   doi = {10.1214/15-STS524},
   issn = {08834237},
   issue = {4},
   journal = {Statistical Science},
   pages = {468-484},
   title = {Functional data analysis of amplitude and phase variation},
   volume = {30},
   year = {2015},
}

@article{Tucker2013,
   author = {J. Derek Tucker and Wei Wu and Anuj Srivastava},
   doi = {10.1016/j.csda.2012.12.001},
   issn = {01679473},
   journal = {Computational Statistics and Data Analysis},
   pages = {50-66},
   publisher = {Elsevier B.V.},
   title = {Generative models for functional data using phase and amplitude separation},
   volume = {61},
   url = {http://dx.doi.org/10.1016/j.csda.2012.12.001},
   year = {2013},
}

@article{Math2005,
author = {Math, Turk J},
pages = {389--401},
title = {{Submanifolds of Riemannian Product Manifolds}},
volume = {29},
year = {2005}
}

@book{Sommer2020,
   author = {Stefan Sommer and Tom Fletcher and Xavier Pennec},
   isbn = {9780128147252},
   title = {Introduction to differential and Riemannian geometry},
   year = {2020},
}

@article{Lang2006,
   author = {Serge Lang},
   issue = {December},
   journal = {Springer},
   pages = {1-6},
   title = {Differential and Riemannian Manifolds},
   volume = {1999},
   year = {2006},
}

@article{Wang1989,
   author = {Tixiang, Wang},
   journal = {Acta Mathematica Sinica},
   pages = {250–262},
   title = {Morse theory on Banach manifolds},
   volume = {5},
   year = {1989},
   url = {https://doi.org/10.1007/BF02107551},
}

@phdthesis{Saba2012,
   author = {Marianna Saba},
   title = {On the usage of the curvature for the comparison of planar curves},
   school = {University of Cagliari}, 
   year = {2012},
}

@article{Surazhsky2002,
   author = {Tatiana Surazhsky and Gershon Elber},
   doi = {10.1142/S0218654302000145},
   issn = {02186543},
   issue = {2},
   journal = {International Journal of Shape Modeling},
   pages = {201-216},
   title = {Metamorphosis of planar parametric curves via curvature interpolation},
   volume = {8},
   year = {2002},
}

@article{Needham2019,
	title = {Shape {Analysis} of {Framed} {Space} {Curves}},
	volume = {61},
	issn = {0924-9907, 1573-7683},
	url = {http://link.springer.com/10.1007/s10851-019-00895-y},
	doi = {10.1007/s10851-019-00895-y},
	language = {en},
	number = {8},
	urldate = {2023-03-04},
	journal = {Journal of Mathematical Imaging and Vision},
	author = {Needham, Tom},
	month = oct,
	year = {2019},
	pages = {1154--1172},
}

@article{Celledoni2016,
author = {Elena Celledoni and Markus Eslitzbichler and Alexander Schmeding},
title = {Shape analysis on Lie groups with applications in computer animation},
journal = {Journal of Geometric Mechanics},
volume = {8},
number = {3},
pages = {273-304},
year = {2016},
doi = {10.3934/jgm.2016008},
}

@article{Lyons2016,
   author = {Robert Lyons},
   journal = {Lecture note},
   title = {Frobenius Theorem Two Ways},
   url = {www.pdx.edu/},
   year = {2016},
}

@book{Wolfgang,
   author = {Wolfgang Kühnel},
   isbn = {9781470423209},
   title = {Differential Geometry Curves – Surfaces},
   volume = {77},
}

@misc{Park2022,
  doi = {10.48550/ARXIV.2203.02398},
  url = {https://arxiv.org/abs/2203.02398},
  author = {Park, Juhyun and Brunel, Nicolas and Chassat, Perrine},
  title = {Curvature and Torsion estimation of 3D functional data: A geometric approach to build the mean shape under the Frenet Serret framework},
  publisher = {arXiv},
  number = {},
  note = {arXiv:2203.02398 [stat]},
  year = {2022},
  copyright = {Creative Commons Attribution Non Commercial No Derivatives 4.0 International}
}

@inproceedings{koestler2022intrinsic,
 author = {Lukas Koestler and Daniel Grittner and Michael Moeller and Daniel Cremers and Zorah Lähner},
 title = {Intrinsic Neural Fields: Learning Functions on Manifolds},
 booktitle = {European Conference on Computer Vision (ECCV)},
 year = {2022},
 eprint = {2203.07967},
 eprinttype = {arXiv},
}

@inproceedings{eisenberger2020hamiltonian,
 author = {Marvin Eisenberger and Daniel Cremers},
 title = {Hamiltonian Dynamics for Real-World Shape Interpolation},
 booktitle = {European Conference on Computer Vision (ECCV)},
 year = {2020},
 award = {Spotlight Presentation},
}

@inproceedings{LRSBC16,
 author = {Zorah Lähner and Emanuele Rodolà and Frank R. Schmidt and Michael M. Bronstein and Daniel Cremers},
 title = {Efficient Globally Optimal 2D-to-3D Deformable Shape Matching},
 booktitle = {IEEE Conference on Computer Vision and Pattern Recognition (CVPR)},
 month = {May},
 year = {2016},
 url = {http://vision.in.tum.de/~laehner/Elastic2D3D/},
}

@book{FlashBerthoz2021,
    Author = {Flash, Tamar and Berthoz, Alain},
     Title = {Space-Time Geometries for Motion and Perception in the Brain and the Arts},
 Publisher = {Springer},
      Year = {2021},
  series={Lecture Notes in Morphogenesis},
}

@article{Pollick2009,
	Author = {Frank E. Pollick and Uri Maoz and Amir A. Handzel and Peter J. Giblin and Guillermo Sapiro and Tamar Flash},
	Journal = {Cortex},
	Pages = {325--339},
	Title = {Three-dimensional arm movements at constant equi-affine speed.},
	Volume = {45},
	Year = {2009}}

@article{Maoz2009,
	Author = {Maoz, Uri and Berthoz, Alain and Flash, Tamar},
	Journal = {Journal of Neurophysiology},
	Pages = {1002--1015},
	Title = {Complex unconstrained three-dimensional hand movement and constant equi-affine speed.},
	Volume = {101},
	Year = {2009}}

@article{Lacquaniti1983,
	Author = {Francesco Lacquaniti and Carlo Terzuolo and Paolo Viviani},
	Journal = {Acta Psychologica (Amst.)},
	Pages = {115--130},
	Title = {The law relating the kinematic and figural aspects of drawing movements.},
	Volume = {54},
	Year = {1983}}

@book{Fan1996,
  Title = {Local polynomial modelling and its applications},
  series =  {CRC monographs on statistics and applied probability 66},
  Author = {Fan, Jianqing and Gijbels, Irene},
  Publisher = {Chapman \& Hall},
  Year = {1996}
}

@article{Sangalli2009,
  title={Efficient estimation of three-dimensional curves and their derivatives by free-knot regression splines, applied to the analysis of inner carotid artery centrelines},
  author={Sangalli, Laura M and Secchi, Piercesare and Vantini, Simone and Veneziani, Alessandro},
  journal={Journal of the Royal Statistical Society: Series C (Applied Statistics)},
  volume={58},
  number={3},
  pages={285--306},
  year={2009},
  publisher={Wiley Online Library}
}
}

\end{document}